\documentclass[aps,pra,10pt,superscriptaddress,twocolumn,floatfix,nofootinbib]{revtex4-2}
\usepackage{xcolor}
\usepackage{graphicx,soul}
\usepackage[normalem]{ulem}

\usepackage{bm}
\usepackage{dsfont}
\usepackage{amsmath, amsthm, amssymb,mathrsfs,dsfont}
\usepackage{xcolor}
\usepackage{setspace}
\usepackage[utf8]{inputenc}
\usepackage[toc,page]{appendix}
\usepackage{hyperref}
\usepackage{mdframed}
\usepackage{footnote}
\usepackage{orcidlink}
\usepackage{booktabs}
\usepackage[capitalise]{cleveref}
\crefname{equation}{Eq.}{Eqs.}
\crefname{figure}{Fig.}{Figs.}
\crefname{section}{Sec.}{Secs.}
\crefname{table}{Table}{Tables}
\crefname{theorem}{Theorem}{Theorems}
\crefname{lemma}{Lemma}{Lemmas}
\crefname{proposition}{Proposition}{Propositions}
\crefname{corollary}{Corollary}{Corollaries}
\crefname{definition}{Definition}{Definitions}
\crefname{appendix}{Appendix}{Appendices}

\newcommand{\ket}[1]{\left| #1 \right\rangle}
\newcommand{\bra}[1]{\left\langle #1 \right|}

\newcommand{\unitvec}[1]{\hat{\bm{#1}}}
\newcommand*{\id}{{\normalfont\hbox{1\kern-0.15em \vrule width .8pt depth-.5pt}}}

\begin{document}

\title{Interaction-Free Polarimetry}

\author{Shushmi Chowdhury \orcidlink{0009-0005-7658-713X}}
\affiliation{Quantum and Light Matter Interaction Theory Group, Department of Physics, University of Strathclyde, Glasgow, G4 0NG, UK}
\author{John Rarity}
\affiliation{Quantum Engineering Technology Laboratories, Department of Electrical and Electronic Engineering, University of Bristol, Woodland Road, Bristol, BS8 1US, UK}
\author{Jonte R. Hance\,\orcidlink{0000-0001-8587-7618}}
\email{jonte.hance@newcastle.ac.uk}
\affiliation{Quantum Group, School of Computing, Newcastle University, 1 Science Square, Newcastle upon Tyne, NE4 5TG, UK}

\begin{abstract}
    The combination of interaction-free measurement and the quantum Zeno effect has been shown to both increase the signal-to-noise ratio of imaging, and reduce the energy absorbed by the imaged object. So far, however, it has not been used for quantitative polarimetry, the reconstruction of a sample's Mueller matrix. In this paper, we extend this to the polarimetry of diattenuating samples. For polarisation components that the sample strongly absorbs, the protocol identifies the sample with far less absorbed energy than a single-pass measurement, a benefit for delicate samples probed with high-frequency radiation. For weakly absorbing components the repeated passes through the sample absorb more, so the benefit is confined to strongly diattenuating features. A Fisher-information analysis shows that, for a sample with a lossless transmitting axis, the protocol locates the absorbing axis with far fewer absorbed photons than single-pass polarimetry, while it measures partial transmittances no more efficiently per absorbed photon, and that the same sorting of lossy components to one output port responds to retardance and depolarisation.
\end{abstract}

\maketitle

\section{Introduction}

Polarimetry is the task of identifying the effect a given thin sample has on the polarisation of light transmitted through that sample \cite{Tyo2006ReviewPolarimetry,Goldberg2020QPolarimetry}. This allows the discrimination of superficially similar chemical samples (e.g. discriminating between different isomers, despite their identical chemical composition \cite{boehme1982spectrophotometric,muller2001isolation,duell2021determination}).
However, as polarimetry requires light flux through the sample, it can easily cause energy-related degradation, especially in samples formed of complex, photosensitive molecules. A way of performing polarimetry with less energy absorbed by the sample would therefore be an improvement to the technique. Quantum correlations already offer one route: Restuccia et al.~measured the optical activity of a chiral solution in a ghost-polarimetry configuration. The sample was illuminated only with unpolarised light at the single-photon level, and the rotation was obtained from coincidences with a polarisation measurement on the partner photon \mbox{\cite{Restuccia2022GhostPolarimetry}}. Related entangled-photon polarimetry of dichroic and birefringent samples at low light has been reported in \mbox{\cite{Magnitskiy2022GhostPolarimetry,Xie2025Anisotropic,Yoon2020QPolSinglePhotons}}. Interaction-free measurement offers a complementary route, in which the absorption by the sample itself is reduced.

Interaction-free measurement involves using single-particle interferometry to detect whether or not there is a block on the arm of an interferometer without the particle travelling down that path. Elitzur and Vaidman initially described this phenomenon in their `bomb detector thought experiment', where there is a chance to see whether a potentially faulty bomb works without detonating it \cite{Elitzur1993Bomb}. Kwiat et al.~then refined this using the quantum Zeno effect, to reduce the chance of the particle going down the arm with the blocker (or bomb) to zero in the limit of many cycles \cite{Kwiat1995IFM}.

Interaction-free measurement has been developed for both classical \mbox{\cite{Salih2013Protocol,Salih2022Laws}} and quantum \mbox{\cite{Salih2016Qubit,Salih2021EFQubit,Salih2020DetTele}} communication (see \mbox{\cite{Hance2021Quantum}} for the foundational status of these protocols). In sensing, it has so far been applied, in proposals and experiments, to imaging opaque and semi-transparent structures \mbox{\cite{White1998IFImaging,Hance2021CFGI,Palici2022IFMMultimode}}, phase objects and polarisation-rotating patterns \mbox{\cite{Zhang2019GhostIFM}}, to measuring partial transparency \mbox{\cite{Jang1999Semitransparent,Mitchison2001Semitransparent,Thomas2014Semitransparency}} and refractive index \mbox{\cite{Liu2020RefractiveIndex}}, and to taking transmission spectra \mbox{\cite{Chen2021IFSpectro}}. Polarisation has appeared either as a degree of freedom of the probe or as a property of the object. In its simplest polarisation form, with the object in the beam inhibiting a stepwise rotation of the polarisation, the Zeno scheme requires a polarising object and detects its presence alone \mbox{\cite{Kwiat1995IFM,Kwiat1999Zeno}}; placing the object in one arm of a polarisation interferometer extends it to any opaque object \mbox{\cite{Kwiat1999Zeno}}, and splitting the probe by polarisation or transverse mode in a single-pass interferometer raises the Elitzur--Vaidman efficiency towards its $50\%$ bound \mbox{\cite{Cruz2026IFMDoF}}. White et al.~proposed polarisation-difference interaction-free images \mbox{\cite{White1998IFImaging}}, and a single-pass interaction-free filter that uses the circular dichroism of an optically pumped atomic vapour to select atom-resonant photons has been demonstrated \mbox{\cite{Wolfgramm2011IFMFilter}}. No quantitative polarimetry, the reconstruction of a sample's Stokes response or Mueller matrix, has been reported with interaction-free or quantum Zeno measurement.

In this paper, we give such a protocol for diattenuating samples, with a full simulation of the Mueller-matrix reconstruction and its physicality. It keeps the benefit of the interaction-free approach, low energy absorption, for the polarisation component the sample absorbs.

Our contribution is therefore not the first polarisation sensitivity in an interaction-free image but the first quantitative polarimetry with one. We show how the two output ports of a Zeno interferometer, measured in the three mutually unbiased polarisation bases, give the Stokes vectors from which a diattenuating sample's Mueller matrix follows (Secs.~\mbox{\ref{Sect:DefPolarimetry}} to \mbox{\ref{Reconstructing Mueller matrix}}). We give the mapping between Mueller matrices and the trace-non-increasing quantum operations that describe absorbing samples, applied to the port-resolved maps of the interferometer, the maps to each output port taken separately (Sec.~\mbox{\ref{sec:quantumrep}} and Appendix~\mbox{\ref{App:MapDenstoStokes}}). We also show where the energy advantage holds, namely for strongly absorbing polarisation components, and where repeated passes through the sample remove it (Sec.~\mbox{\ref{sec:disc}}). A Fisher-information analysis (Sec.~\mbox{\ref{sec:fisher}}) shows that the information advantage lies in locating the absorbing axis of a sample with a lossless axis rather than in measuring partial transmittances per absorbed photon, and Sec.~\mbox{\ref{sec:beyond}} shows how retardance and depolarisation enter the same sorting.

\section{Defining Polarimetry}\label{Sect:DefPolarimetry}

\begin{figure}
\includegraphics[width=1\linewidth]{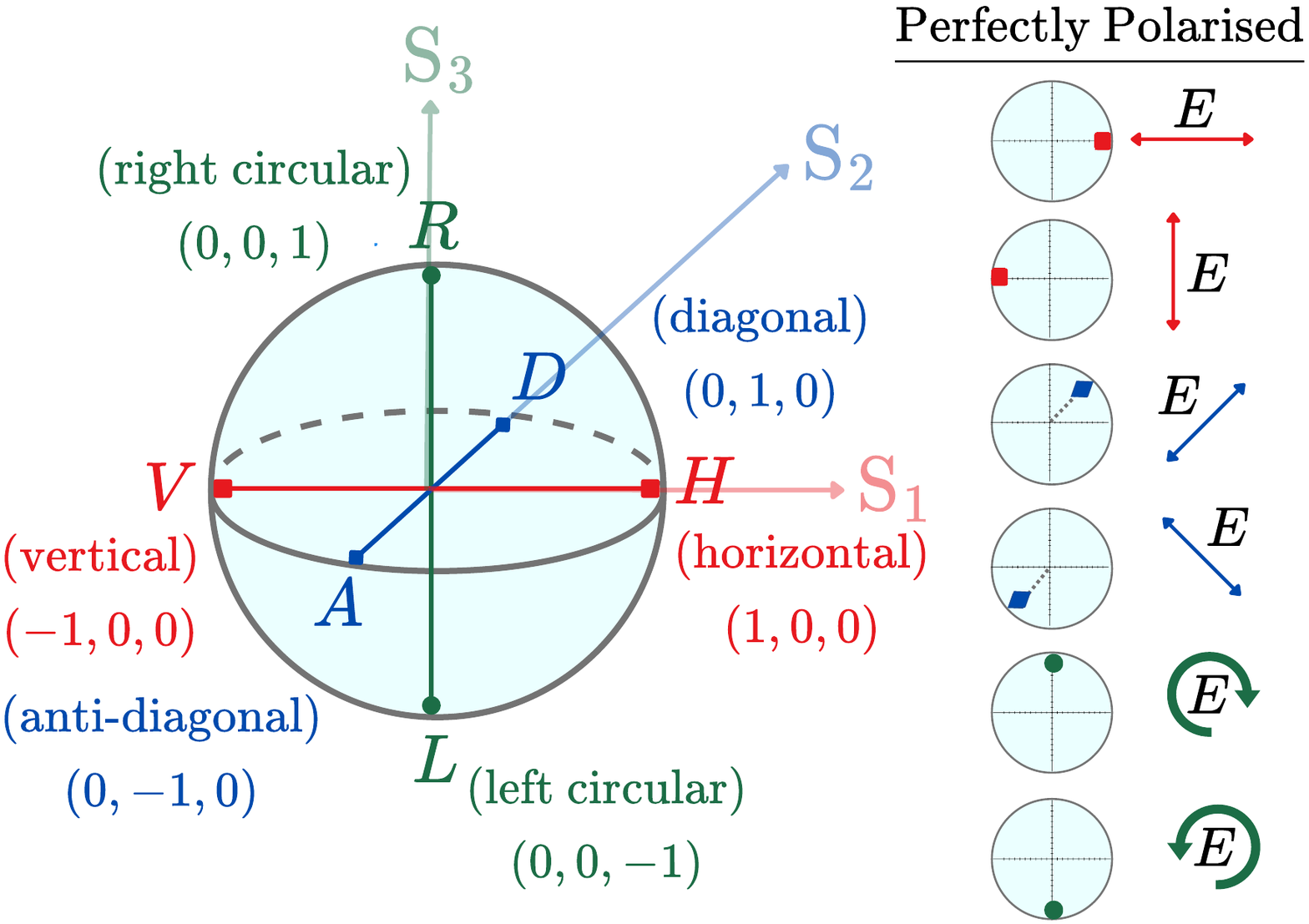}
\caption{\label{fig: spere}  Schematic showing the Poincar\'e sphere representing the states of perfectly polarised light using the normalised Stokes vector $\mathbf{S}=(S_1,S_2,S_3)/S_0$. The six canonical pure polarisation states, horizontal (H), vertical (V), diagonal (D), anti-diagonal (A), right circular (R), and left circular (L), are shown with their corresponding electric field orientations.
}
\end{figure}

This section follows the summary of classical polarimetry in Ref.~\mbox{\cite{Goldberg2020QPolarimetry}}, Sec.~I (see also \mbox{\cite{Tyo2006ReviewPolarimetry}}); the retarder-diattenuator-depolariser decomposition is that of Lu and Chipman \mbox{\cite{Lu1996PolarDecomposition}}.
Classically, we treat a polarised monochromatic light wave travelling in direction $\mathbf{k}$ as
\begin{equation}
    \mathbf{E}(\mathbf{r},t)=(h \unitvec{h}+v \unitvec{v})e^{i\mathbf{k}\cdot\mathbf{r}-i\omega t}
\end{equation}
where $h$ and $v$ are complex constants, and $\unitvec{h}$ and $\unitvec{v}$ are unit vectors orthogonal to both each other and $\mathbf{k}$. We categorise the polarisation properties of this wave $\mathbf{E}$ using the Stokes parameters \cite{Goldberg2020QPolarimetry}

\begin{equation}
    \begin{split}
        &S_0 = |\unitvec{h} \cdot \mathbf{E}|^2 + |\unitvec{v} \cdot \mathbf{E}|^2 = |h|^2 + |v|^2\\
        &S_1 = |\unitvec{h} \cdot \mathbf{E}|^2 - |\unitvec{v} \cdot \mathbf{E}|^2 = |h|^2 - |v|^2\\
        &S_2 = \text{2 Re}[(\unitvec{h} \cdot \mathbf{E})^*\cdot(\unitvec{v} \cdot \mathbf{E})] = h^* v + h v^*\\
        &S_3 = \text{2 Im}[(\unitvec{h} \cdot \mathbf{E})^*\cdot(\unitvec{v} \cdot \mathbf{E})] = -i(h^* v - h v^*).
    \end{split}
\end{equation}
For this fully polarised wave, $S_0^2 = S_1^2 + S_2^2 + S_3^2$, meaning we can define a vector $\mathbf{S}\equiv(S_1,S_2,S_3)$, which, when normalised by $S_0$, lies on the surface of the unit Poincar\'e sphere. In this way, we can associate $S_1$ with the $H/V$ basis, $S_2$ with the $D/A$ basis, and $S_3$ with the $R/L$ basis, shown in Fig.~\ref{fig: spere}.

When dealing with stochastic (partially polarised) light, this vector $\mathbf{S}/S_0$ lies inside the Poincar\'e sphere, and its length gives the degree of polarisation:
\begin{equation}P_{\mathrm{pol}}=|\mathbf{S}|/S_0.
\end{equation}

All classical light rays can be decomposed into $P_{\mathrm{pol}}$ times a perfectly polarised ray, plus $(1-P_{\mathrm{pol}})$ times a perfectly unpolarised ray. Using this, we can now formally define polarimetry, as identifying how a material linearly transforms the Stokes parameters of an incoming light ray. For a given material, this transform will be by the $4\times 4$ Mueller matrix $M$, where

\begin{equation}
    S_\mu \rightarrow \sum_{\nu=0}^3 M_{\mu,\nu}S_\nu .
\end{equation}

Therefore, the goal of polarimetry is to identify the components of the Mueller matrix of our object. This is typically done classically by shining bright, perfectly polarised light onto the object, repeating this for different polarisation orientations (to cover the two orthogonal polarisation states of each of the three mutually unbiased bases (MUBs) for polarisation, i.e. $H/V$, $D/A$ and $R/L$), and then using matrix inversion.

For this, we require the output Stokes parameters, which can be obtained from the intensities of light output after different polarisers:
\begin{equation}
    \begin{split}
        S_0 = I_H + I_V\\
        S_1 = I_H - I_V\\
        S_2 = I_D - I_A\\
        S_3 = I_R - I_L,\\
    \end{split}
\end{equation}
where $I_X$ is the intensity of the output beam after it has been completely polarised along a given direction, $X$. Using the 36 parameters we get (6 output parameters for each of the 6 input polarisations), we can obtain the Mueller matrix for any generalised optically-transmissive medium.

In this paper, we aim to replace this bright polarised light with polarised single photons, so as to make use of interaction-free measurement and the quantum Zeno effect of light being absorbed by the object under investigation.

Any Mueller matrix can be decomposed into the product of a diattenuator, a retarder and a depolariser, uniquely for a given ordering of the three \mbox{\cite{Lu1996PolarDecomposition}}.

A diattenuator transmits the two polarisations along its principal axes with different intensities, the principal transmittances $T_{\max}$ and $T_{\min}$. It therefore maps perfectly polarised light to perfectly polarised light of a different intensity, and can change the degree of polarisation of partially polarised light \mbox{\cite{Lu1996PolarDecomposition}}.

The generalised form for Mueller matrices for pure diattenuators is

\begin{equation}
M_D = \tau
\begin{pmatrix}
1 & \mathbf{D}^{T}\\
\mathbf{D} & \mathbf{m}_D
\end{pmatrix},
\end{equation}
where
\begin{equation}
\begin{split}
\mathbf{D} &=
\begin{pmatrix}
D_{H/V}\\
D_{D/A}\\
D_{R/L}
\end{pmatrix}, \qquad
D_{\mathrm{MUB}} = \frac{I_1-I_2}{I_1+I_2},\\[6pt]
\hat{\mathbf{D}} &=
\frac{\mathbf{D}}{\lVert\mathbf{D}\rVert},
\qquad \lVert\mathbf{D}\rVert>0,\\[6pt]
\mathbf{m}_D &=
\sqrt{1-\lVert\mathbf{D}\rVert^2}\,\id_3
+
\left(1-\sqrt{1-\lVert\mathbf{D}\rVert^2}\right)
\hat{\mathbf{D}}\hat{\mathbf{D}}^{T},
\end{split}
\end{equation}
$I_1$ and $I_2$ are the intensities measured from the two orthogonal input polarisation states of the input MUB with equal incident intensities in both measurements, and $\tau$ is the intensity transmittance for unpolarised input light. We note for $\mathbf{D}=\mathbf{0}$, $\mathbf{m}_D=\id_3$.

The above matrix can be extended to a generalised form to include retardance
and depolarisation using the Lu--Chipman
decomposition~\mbox{\cite{Lu1996PolarDecomposition}},
\[
M=M_{\Delta}M_RM_D,
\]
where $M_D$ is the pure diattenuator matrix defined above, and the retarder and depolariser matrices are, respectively,
\[
M_R=
\begin{pmatrix}
1 & \mathbf{0}^{T}\\
\mathbf{0} & \mathbf{R}
\end{pmatrix},
\qquad
M_{\Delta}=
\begin{pmatrix}
1 & \mathbf{0}^{T}\\
\mathbf{P}_{\Delta} & \mathbf{m}_{\Delta}
\end{pmatrix}.
\]

Here, $\mathbf{R}$ represents a $3\times3$ submatrix of $M_R$ acting on the polarisation components of the Stokes vector and introduces a relative phase shift $\delta$ 
between two orthogonal components of the retarder, $\mathbf{P}_{\Delta}$ is the polarizance vector of $M_\Delta$ and $\mathbf{m}_{\Delta}$ is its
real symmetric $3\times3$ depolarising submatrix. The matrices do not commute so order of application is important.

In this paper we focus on the polarimetry of diattenuating samples, because the interaction-free suppression of absorption acts on the polarisation component a sample absorbs (Sec.~\mbox{\ref{sec:disc}}); retarding and depolarising samples are taken up in Sec.~\mbox{\ref{sec:beyond}}.

\section{Quantum Representation of \texorpdfstring{ Polarisation}{Polarisation} Transformations}\label{sec:quantumrep}

The polarisation state of a photon in the \(H/V\) basis can be expressed in terms of the Stokes parameters through the density matrix \cite{nielsen_chuang_2010,PhysRevA.87.064102}, (for details, see Appendix~\ref{App:MapDenstoStokes})
\begin{equation}
    \rho=\frac{1}{2}\left(S_0I+S_1\sigma_z+S_2\sigma_x+S_3\sigma_y\right),
    \label{dens_stokes}
\end{equation}
where $I$ denotes the identity matrix and $\sigma_x$, $\sigma_y$, and $\sigma_z$ are the Pauli matrices. 

Conversely, in this basis, the Stokes parameters can be extracted using the inverse mapping by multiplying both sides of Eq.~\ref{dens_stokes} by $\alpha_k$ using the linearity of the trace (for details, see Appendix~\ref{App:MapDenstoStokes}), 
\begin{equation}
    S_k = \text{Tr}(\rho \alpha_k),
\end{equation}
where the basis operators are defined as  $(\alpha_0,\alpha_1,\alpha_2,\alpha_3)=(I, \sigma_z, \sigma_x, \sigma_y)$ and satisfy the trace orthogonal condition $\text{Tr}(\alpha_j \alpha_k) = 2\delta_{jk}$.  
For a normalised state, $S_0=1$ in the above equations. For unnormalised output at a specific detector, $S_0$ represents the probability of the photon reaching that port, given that input probability is unity. 

On the other hand, any physical linear transformation of a density matrix can be described by a quantum operation $\mathcal{I}$, a quantum channel when it preserves the trace; here it need not, since the sample absorbs \mbox{\cite{nielsen_chuang_2010}}. Acting with $\mathcal{I}$ on $\rho$ from Eq.~\mbox{\eqref{dens_stokes}} and using its linearity gives the output state, 
$\rho' = \frac{1}{2}\sum_{k=0}^{3}S_k\mathcal{I}(\alpha_k)$. To compute the output Stokes parameters, we can do the same inverse mapping as before using the linearity of the trace giving $S'_j=\frac{1}{2}\sum_{k=0}^{3} S_k\text{Tr}\big( \alpha_j\mathcal{I}(\alpha_k)\big)$. Now using the relation $S' = MS$, we can express the Mueller matrix $M$ as \mbox{\cite{GOLDBERG2022185,Aiello2007LinOptQMaps}} (for details, see Appendix~\ref{App:MuellerfromChannel})
\begin{equation}
    M_{jk} = \frac{1}{2}\text{Tr}\big( \alpha_j\mathcal{I}(\alpha_k)\big),
    \label{eq:muell-chann}
\end{equation}
where the physical action of the quantum operation $\mathcal{I}$ can be defined by Kraus operators:
\begin{equation}
    \mathcal{I}(\alpha_k) = \sum_\ell E_\ell \alpha_k E_\ell^\dagger,
\end{equation}
with $\sum_\ell E_\ell^\dagger E_\ell \leq I$ for a trace-non-increasing operation (the inequality allows for absorption).

We can also consider the inverse problem of reconstructing the action of the corresponding quantum map from a given Mueller matrix $M$. Because the Pauli operators $(\alpha_0,\alpha_1,\alpha_2,\alpha_3)=(I,\sigma_z,\sigma_x,\sigma_y)$ constitute a complete orthogonal basis under the Hilbert--Schmidt inner product, a linear map is uniquely determined by its action on these four operators. Since the $k$-th column of the Mueller matrix provides the Pauli-basis coefficients of $\mathcal I(\alpha_k)$, the inverse mapping from the Mueller matrix to the quantum map is given by (for details, see Appendix~\ref{App:Deriving_Quantum Channel_Mueller})
\begin{equation}
\mathcal I_M(\alpha_k)
=
\sum_{j=0}^{3}
M_{jk}\alpha_j.
\label{eq:channel_from_mueller1}
\end{equation}

While Eq.~\ref{eq:channel_from_mueller1} uniquely determines the linear action of the reconstructed map on a density matrix, this reconstruction only holds as a valid physical process when the resulting operation satisfies complete positivity ($C_{\mathcal I} \ge 0$) and the trace bounds ($0 \le T \le I$) \mbox{\cite{Aiello2007LinOptQMaps,Simon2010NonquantumEntanglement}}, where $C_{\mathcal I}$ is the Choi matrix representing the map and $T = \operatorname{Tr}_{\mathrm{out}}(C_{\mathcal I})$ is its partial trace taken over the output Hilbert space (for details, see Appendix~\ref{Physical_val}).

\section{Protocol}

\begin{figure*}[ht]
    \centering
    \includegraphics[width=\linewidth]{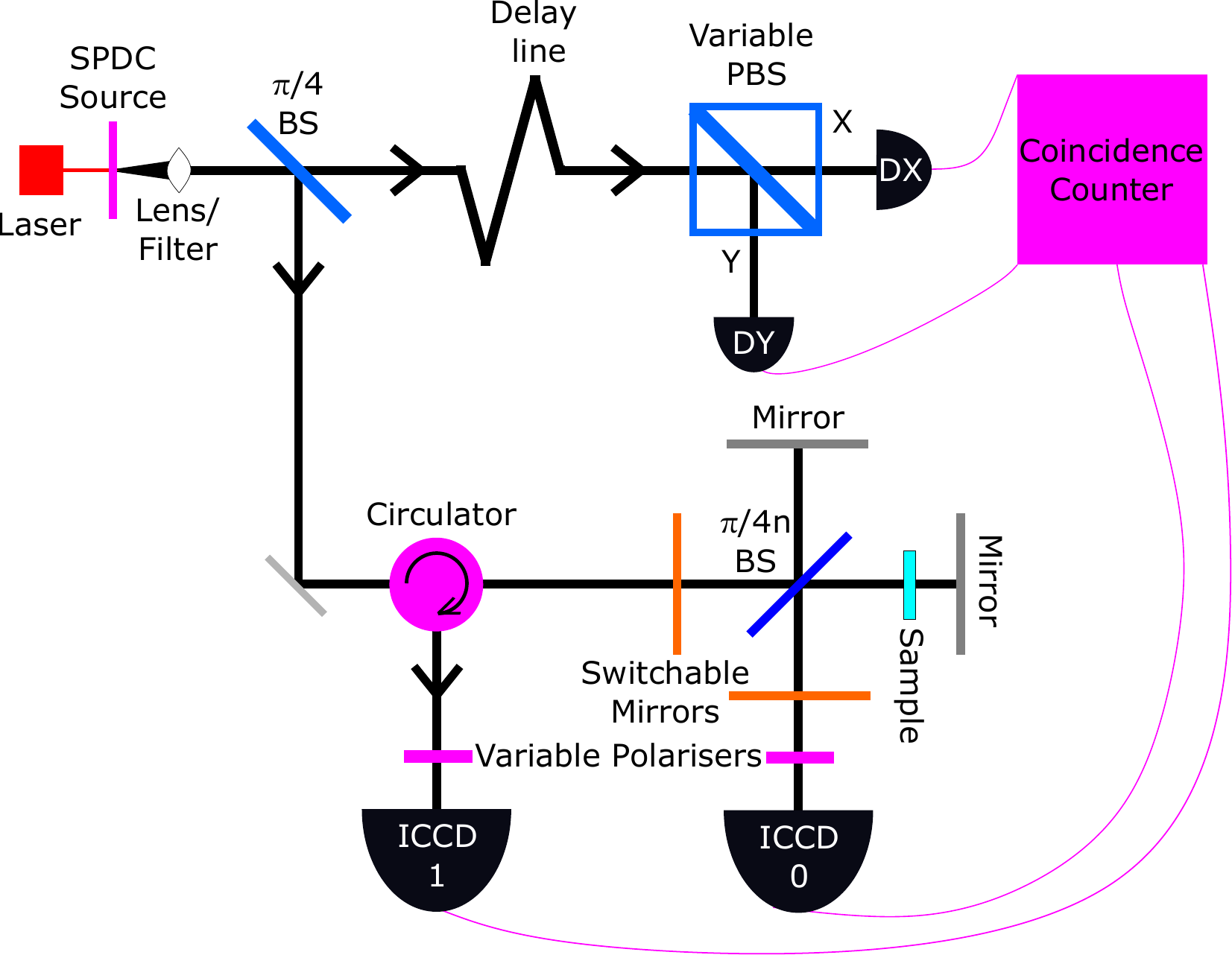}
    \caption{Protocol for using an entangled photon pair to image a transparent, potentially polarising object, using interaction-free measurement and the quantum Zeno effect to reduce the energy absorbed by the object. The variable polarising beam splitter (PBS) allows us to pick a polarisation basis to image in ($H/V$, $D/A$, or $R/L$), by using wave plates to rotate from that basis into the $H/V$ basis, then applying a standard $H/V$ polarising beam splitter. The switchable mirrors allow the photon into the interferometer, and then keep it there for $n$ cycles. The $\pi/4n$ beam splitter is mostly reflecting (reflection amplitude $\cos(\pi/4n)$), so the photon stays mostly in the circulator port and the empty arm; only an amplitude $\sin(\pi/4n)$ leaks into the sample arm on each outgoing pass. A photon blocked by the sample therefore exits through the circulator to ICCD1, and an unblocked photon through the lower port to ICCD0. Coincidences between the ICCDs and the heralding detectors DX and DY give the polarisation-resolved images. As the number of cycles $n$ goes to infinity, the probability of the photon being absorbed by an opaque region of the object goes to 0 (as $\pi^2/4n$), while a transparent region is traversed on every cycle.}
    \label{fig:PolznImg}
\end{figure*}

To achieve interaction-free polarimetry, we start with an apparatus similar to that used to counterfactually ghost image objects in \cite{Hance2021CFGI}. However, it is arranged both to allow the probing of an object with different polarisations of light, and so the high-resolution intensified charge-coupled device (ICCD) cameras receive the photon from the interaction-free measurement protocol directly, with what was the ghost-path in \cite{Hance2021CFGI} now being used for polarisation-based heralding. We give an apparatus for this in Fig.~\ref{fig:PolznImg}.

In this protocol, we use a polarisation and position-momentum entangled pair of photons, split at our $\pi/4$ (50:50) beam splitter. We send one photon to a multi-pixel version of the high-efficiency (quantum Zeno) interaction-free measurement scheme of Kwiat et al.~\mbox{\cite{Kwiat1995IFM,Kwiat1999Zeno}}, here in Michelson form (Fig.~\mbox{\ref{fig:PolznImg}}), with the object to be imaged in one path (the imaging photon). The other photon goes to our variable polarising beam splitter (PBS), where we can choose a polarisation basis to image the object in (the herald photon). 

On exiting the $\pi/4$ beam splitter, the imaging photon enters this multi-pixel Zeno interferometer, designed to maintain pixel coherence. Here, it enters a $\pi/4n$ beam splitter, where we define a $\theta$ beam splitter as performing a Pauli-Y rotation between two path modes of rotation angle $2\theta$:
\begin{equation}
    \hat{\textbf{R}}_y(2\theta) =
        \begin{pmatrix}
    \cos{(\theta)}  &  -\sin{(\theta)}\\
      \sin{(\theta)}  &  \cos{(\theta)}
        \end{pmatrix}
        = e^{-i\theta\hat{\sigma}_y}.
        \label{eq.beam}
\end{equation}

This sends some probability amplitude of the photon to the object (where it is either absorbed, or travels through it, to a mirror, back through it, and back to the beam splitter, or some combination of the two), and most of it via an empty path to a mirror and back to the $\pi/4n$ beam splitter. These two paths are the same length, to ensure no phase difference between the two components. They interfere at the beam splitter, some parts reflecting and some transmitting, and on both transmitted and reflected path are reflected off the switchable mirrors. Therefore, these components return to again interfere at the $\pi/4n$ beam splitter. This constitutes one cycle. After $n$ cycles, the switchable mirrors go from reflective to transmissive, and so the photon, if not absorbed, travels to either ICCD0 or ICCD1 depending on interference.

These two outputs correspond to the photon being blocked (ICCD1) or not blocked (ICCD0) by the object, and have variable polarisers placed in front of them, to allow us to choose a detection polarisation ($H$, $V$, $D$, $A$, $R$ or $L$).

The variable-PBS (heralding) arm uses a  quarter-wave, then half-wave, then another quarter-wave plate to rotate the photon from the imaging basis (either $H/V$, $D/A$ or $R/L$) into the $H/V$ basis. We then split the two orthogonal polarisations for separate detection using a normal ($H/V$) PBS. We record the correlations between either of the two ICCDs and either of the two polarisation-basis detectors ($DX$ and $DY$). By building up correlation-based images of the object in each of the three mutually unbiased bases (MUBs) which exist for polarisation (i.e. $H/V$, $D/A$, $R/L$), we can slowly build up polarisation statistics for the object, allowing us to derive the Mueller matrices for each of the pixel-regions of the object.

Unless $n\to\infty$, a polarisation component that the sample absorbs is still absorbed with a small probability (about $\pi^2/4n$ for an opaque component), so the returning intensity is reduced in a polarisation-dependent way. This dependence is the signal: the protocol compares the two output ports and the absorbed fraction for each input polarisation (Sec.~\mbox{\ref{sec:results}}).

\subsection{Numerical Method}\label{sec:numerical}
The spontaneous parametric down-conversion (SPDC) source generates a maximally entangled polarisation state:
\begin{equation}
    |\Phi^+\rangle_{hi}
    =
    \frac{|H\rangle_h|H\rangle_i
    +|V\rangle_h|V\rangle_i}{\sqrt{2}},
\end{equation}
where subscripts $h$ and $i$ denote the herald and imaging photons, respectively, with the density matrix of the state being $\rho_{hi}=|\Phi^+\rangle\langle\Phi^+|$.

Measuring the herald photon prepares a corresponding polarisation state for the imaging photon. If the herald is detected in state \( |\chi\rangle_h \), the imaging photon is left in the conditional state
\begin{equation}
\rho_{i|\chi}
=
\frac{
\operatorname{Tr}_h\!\left[
(\Pi_\chi\otimes I_{\mathrm{pol}})
\rho_{hi}
(\Pi_\chi\otimes I_{\mathrm{pol}})
\right]
}{
p_\chi
},
\end{equation}
where $\Pi_\chi=|\chi\rangle\langle\chi|$ is the projector onto one of the six polarisation states and $p_\chi=\operatorname{Tr}\!\left[(\Pi_\chi\otimes I_{\mathrm{pol}})\rho_{hi} \right]$ is the probability of obtaining that herald result. For linear states ($H, V, D, A$), herald measurements result in the same polarisation state in the imaging arm. For circular polarisation, projecting an $L$ herald state leaves the corresponding imaging photon in $R$ and likewise, projection of herald state $R$ results in an $L$ imaging photon.
The imaging photon travels via path 2 to the interferometer, shown in Fig.~\ref{fig:PolznImg}, and the joint input state is
$ \rho_{\mathrm{in}}= P_2\otimes\rho_{i|\chi}.$ The imaging photon has a spatial degree of freedom, spanned by $\{|1\rangle,|2\rangle\}$, and a polarisation degree of freedom,
spanned by $\{|H\rangle,|V\rangle\}$.  The total Hilbert space is therefore
$\mathcal{H}=\mathcal{H}_{\mathrm{path}} \otimes\mathcal{H}_{\mathrm{pol}}$
with ordered basis $\{|1,H\rangle,|1,V\rangle,|2,H\rangle,|2,V\rangle\}$.
For our simulation we put the sample in path 1 with corresponding projector, $ P_1=|1\rangle\langle1|$ and path 2 is the empty unblocked path with projector $P_2=|2\rangle\langle2|$.  The photon then encounters the $\pi/4n$ beam splitter in Eq.~\ref{eq.beam} in the interferometer, the joint operation in the spatial and polarisation spaces of which can be expressed as 
\begin{equation}
    U_{\mathrm{BS}} = \hat{\textbf{R}}_y(2\theta_B)\otimes I_{\mathrm{pol}},
\end{equation}
with $\theta_B=\pi/4n$.
The beam splitter acts on the photon's spatial degrees of freedom while leaving the polarisation unchanged as denoted by the identity operator.
There are two beam splitter encounters in each physical cycle.

\subsubsection{Modelling the sample}
\begin{figure*}[t]
    \centering
    \includegraphics[width=\textwidth]{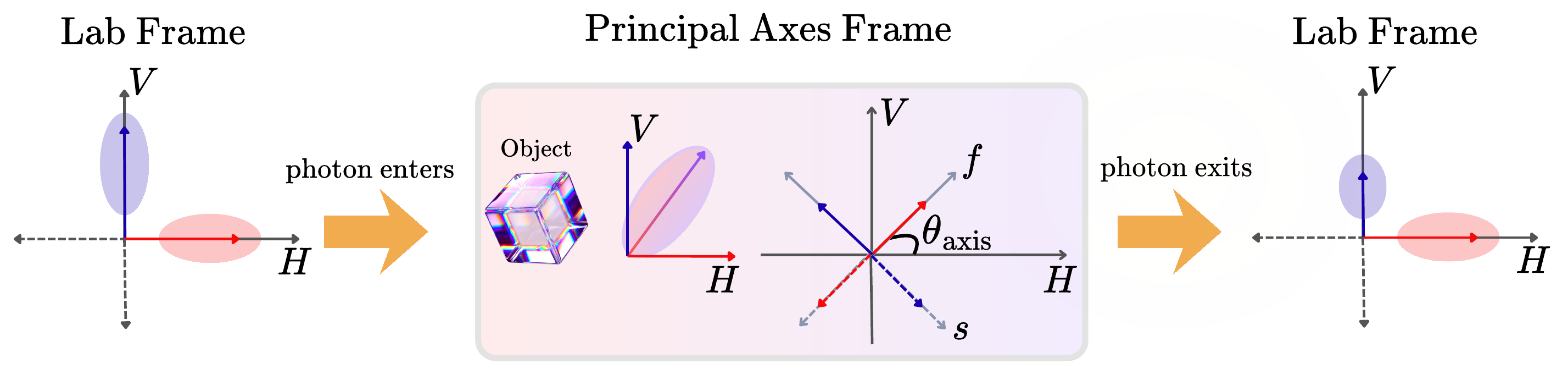}
    \caption{Schematic showing the interaction of a photon with the object. 
The laboratory basis is defined by the horizontal and vertical polarisation components, $H$ and $V$.
The optical object has its own principal axes, denoted $f$ with principal transmittance $T_f$, and $s$ with principal transmittance $T_s$. These are rotated by an angle $\theta$ with respect to the laboratory $H$ axis. Upon entering the object, the photon's polarisation components are expressed in the $f/s$ basis and the resulting state is then transformed back into the laboratory $H/V$ basis for subsequent detection at the ICCDs.
}
\label{fig:diatt}
\end{figure*}

For the simple case considered in this work, we model the sample as a rotated linear diattenuator using Jones calculus. In Fig.~\ref{fig:diatt}, we show a schematic of a photon's interaction with a diattenuating sample. To show the effect of the sample on the photon's polarisation state after a single pass through the sample, we construct a Jones matrix from the optical properties of the sample as 
\begin{equation}
\begin{split}
    J(\theta_{\mathrm{axis}},T_f,T_s,\delta)
    &= R(\theta_{\mathrm{axis}})
    \begin{pmatrix}
        \sqrt{T_f} & 0\\
        0 & \sqrt{T_s}\,e^{i\delta}
    \end{pmatrix}\\
    &\quad\times R^{\dagger}(\theta_{\mathrm{axis}}),
\end{split}
\end{equation}

where $T_f$ and $T_s$ are the principal transmittances along the principal axes $f$ and $s$ (the fast and slow axes even when $\delta\neq0$), $\delta$ is relative retardance, and $R(\theta)=\hat{\textbf{R}}_y(2\theta)$ is the rotation matrix of Eq.~\mbox{\eqref{eq.beam}}, so that the fast axis lies at angle $\theta_{\mathrm{axis}}$ from $H$ (with the order $R^{\dagger}\,\mathrm{diag}\,R$ it would lie at $-\theta_{\mathrm{axis}}$, which is not the convention of Fig.~\mbox{\ref{fig.det-abs}}). The pure diattenuator simulations use $\delta=0$.

Interacting with the sample introduces a path-dependent operator, defined as

\begin{equation}
    U_{\mathrm{sample}}
    =
    P_1\otimes J
    +P_2\otimes I_{\mathrm{pol}},
\end{equation}
where $J$ represents the sample in path 1 and the no-sample case is represented by the identity operator in path 2. $J$ is non-unitary when the sample is lossy and thus the trace of $\rho\mapsto U_{\mathrm{sample}}\rho U_{\mathrm{sample}}^{\dagger}$ is not preserved.

During each cycle, the photon encounters the sample twice in its journey on the outward and return paths. Assuming ideal mirrors, the operator for one cycle, i.e.  beam splitter $\mapsto$ sample $\mapsto$ sample $\mapsto$ beam splitter sequence, is
therefore,
\begin{equation}
    K=U_{\mathrm{BS}}U_{\mathrm{sample}}^2U_{\mathrm{BS}}.
\end{equation}

The output state, after $n$ complete cycles, is thus
\begin{equation}
    \rho_{\mathrm{out}} = K^n\rho_{\mathrm{in}}(K^{\dagger})^n,
\end{equation}
with total survival and absorption probabilities, conditioned on heralding,
given by $P_{\mathrm{surv}} =\operatorname{Tr}(\rho_{\mathrm{out}})$, and $P_{\mathrm{abs}}=1-P_{\mathrm{surv}}$. For an opaque component (an input polarised along an axis with zero transmittance), $K^n$ can be evaluated in closed form. The amplitude remaining in the interferometer is $\cos(\pi/4n)\cos^{n-1}(\pi/2n)$, so $P_{\mathrm{abs}}=1-\cos^2(\pi/4n)\cos^{2(n-1)}(\pi/2n)$. This equals $1/2$ at $n=1$, peaks at $0.573$ for $n=2$ and falls as $\pi^2/4n$ for large $n$ ($0.024$ at $n=100$).

\subsubsection{Polarisation-resolved detection}
Upon reaching the variable detector for complete cycles $n$, we perform polarisation-resolved detection.  In our simulation we assign path~1 to ICCD0 and path~2 to ICCD1. For the case with no sample present, the photon exits through ICCD0 as complete cycles $n$ produce a total path rotation of $\pi/2$ that swaps photon paths from path~2 to path~1, with unit probability for ideal components, as confirmed in Appendix~\mbox{\ref{App:NoSample}}. For a polarisation component that the sample absorbs, the Zeno effect suppresses the transition of the photon's probability amplitude into the sample arm. A transmitted component is not held back, so the photon may exit through ICCD0 or ICCD1, depending on the input polarisation of the light and the sample properties.
Given we have two paths that lead to the detectors $k\in\{1,2\}$, and six standard polarisation states $|\alpha\rangle \in \{|H\rangle,|V\rangle,|D\rangle,|A\rangle,|R\rangle,|L\rangle\}$, 
the joint projector on the output state is
\begin{equation}
    \Pi_{k,\alpha}
    =
    P_k\otimes|\alpha\rangle\langle\alpha|,
\end{equation}
with corresponding detected intensity $I_{k,\alpha}=\operatorname{Tr}\!\left(\Pi_{k,\alpha}\rho_{\mathrm{out}}\right).$
These six analyser measurements are combined to obtain an unnormalised Stokes vector at each detector:
\begin{equation}
    \mathbf{S}^{(k)}_{\mathrm{out}}
    =
    \begin{pmatrix}
        I_{k,H}+I_{k,V}\\
        I_{k,H}-I_{k,V}\\
        I_{k,D}-I_{k,A}\\
        I_{k,R}-I_{k,L}
    \end{pmatrix}.
    \label{e}
\end{equation}

\subsubsection{Reconstructing \texorpdfstring{the}{the} Mueller matrix}
\label{Reconstructing Mueller matrix}
We use the six inputs $H$, $V$, $D$, $A$, $R$, and $L$ for the imaging photon set by the herald, which are converted into density matrices, using Eq.~\ref{dens_stokes} and propagated through the interferometer. At each output port, the final density matrix is projected onto the six polarisation states $H$, $V$, $D$, $A$, $R$, and $L$, and the corresponding intensities are obtained from the traces of the projected states.  These intensities are then used to reconstruct the output Stokes vectors from Eq.~\ref{e} at the two output ICCD ports. We construct a single $4\times1$ output Stokes vector at each port. To characterise the full system, we consider all the input states forming a $4\times6$ matrix $\mathsf{S}_{\mathrm{in}}$, which produces the corresponding $4\times6$ matrix $\mathsf{S}^{(k)}_{\mathrm{out}}$ of measured output Stokes vectors at each ICCD port $k$. The rows represent the four Stokes parameters $S_0, S_1, S_2, S_3$ calculated from the intensities of the six inputs, and each column represents the resulting Stokes vector for one of the six input polarisation states $(H, V, D, A, R, L)$.

The effective Mueller matrix at each port is then reconstructed by employing the Moore--Penrose pseudoinverse:

\begin{equation}
    M_k
    =
    \mathsf{S}^{(k)}_{\mathrm{out}}
    \mathsf{S}_{\mathrm{in}}^{+},
\end{equation}
where $\mathsf{S}_{\mathrm{in}}^{+}$ is a $6\times 4$ Moore--Penrose pseudoinverse of $\mathsf{S}_{\mathrm{in}}$ and the effective map for all surviving photons is $ M_{\mathrm{surv}}=M_{\mathrm{ICCD0}}+M_{\mathrm{ICCD1}}$. Since $\mathsf{S}_{\mathrm{in}}$ has rank 4, the pseudoinverse recovers $M_k$ in the noiseless simulation. For an ideal polariser at large $n$, $M_{\mathrm{ICCD0}}$ is the polariser's own Mueller matrix to within $0.01$ in every element at $n=100$, because the transmitted component rotates fully into the sample arm, while $M_{\mathrm{ICCD1}}$ is $P_{\mathrm{surv}}$ times the Mueller matrix of the orthogonal polariser.

For simulating a diattenuating sample, we use known reference Mueller matrices. For each reference diattenuator, the known values of $(T_f,T_s,\theta_{\mathrm{axis}})$ are first used to simulate the protocol and obtain the corresponding Stokes vectors at the two output ports for the six input states, and the Mueller matrix is obtained using the inverse. These simulated responses serve as the reference measurements. The simulation is then repeated using trial values of the sample parameters, and the resulting output Stokes vectors are compared with the reference responses. Optimisation with least-squares varies $(T_f,T_s,\theta_{\mathrm{axis}})$ until it minimises the difference between the trial output Stokes parameters and those generated using the reference sample parameters. The best-fit parameters are then used to reconstruct the Jones matrix of the sample, from which the corresponding Mueller matrix is obtained using Eq.(\ref{eq:muell-chann}). The parametrisation has a two-fold degeneracy, $(T_f,T_s,\theta_{\mathrm{axis}})$ and $(T_s,T_f,\theta_{\mathrm{axis}}\pm\pi/2)$ describing the same sample, so the fitted parameters are compared through the Mueller matrix rather than directly. For a deterministic, non-depolarising optical element described by a Jones matrix \(J\), the action of the quantum operation \(\mathcal{I}\) is described by 
$\mathcal{I}(\alpha_k)=J\alpha_kJ^\dagger$.
For a lossy element, the Jones matrix satisfies $J^\dagger J \leq I$.

As a check of physicality, we check that the Choi matrix of the reconstructed Mueller matrix  is positive semidefinite and that its partial trace over the output satisfies $\operatorname{Tr}_{\mathrm{out}}(C)\preceq I$, within numerical tolerance. These conditions ensure complete positivity and trace-non-increasing evolution, respectively, allowing sample absorption to be physical (for details, see Appendix~\ref{Physical_val}).

For reference samples, the reconstructed Mueller matrix can then be compared directly with the known matrix using the Frobenius error to assess the accuracy of the recovery procedure.

\begin{figure}
    \centering
    \includegraphics[width=0.92\linewidth]{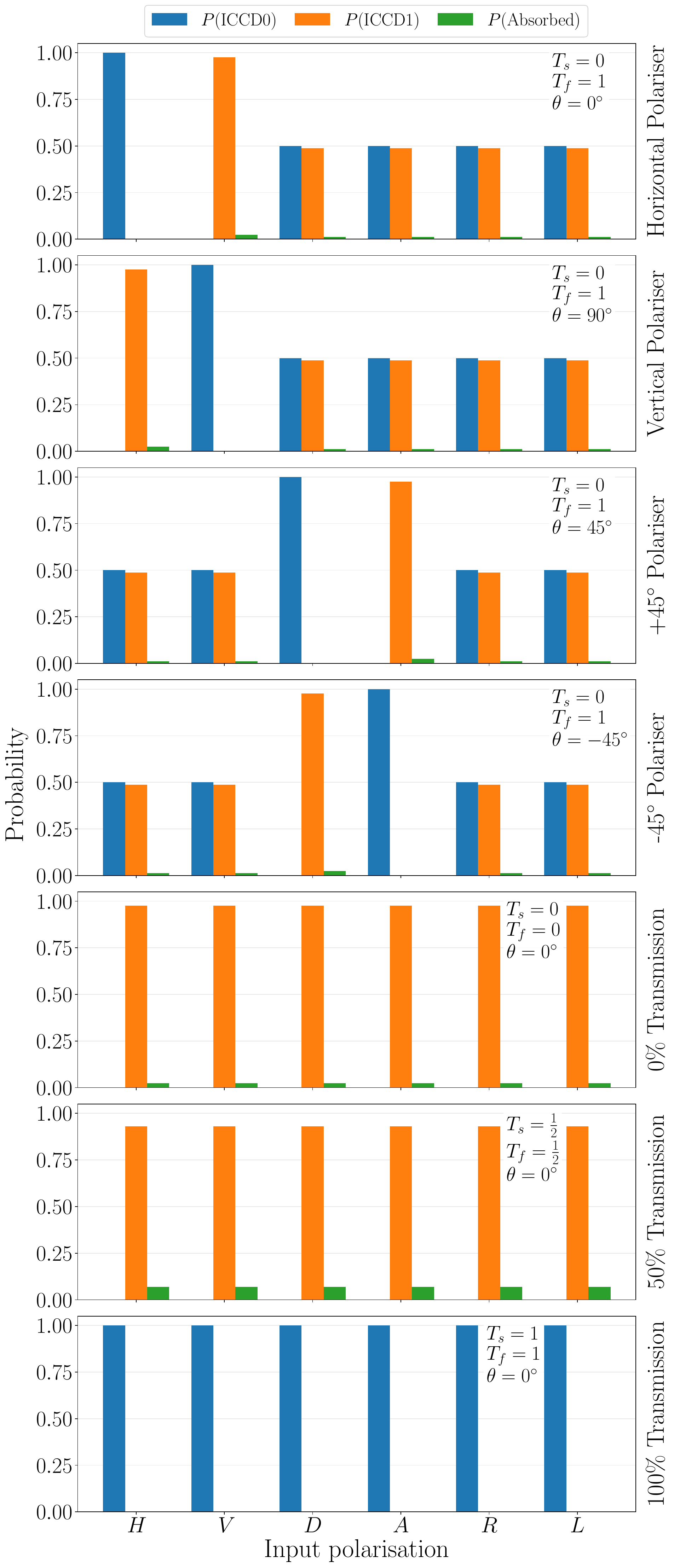}
    \caption{Probabilities per incident photon of detecting the imaging photon at ICCD0 (blue) and ICCD1 (orange), together with the probability of absorption by the sample (green), after $n=100$ interferometer cycles. The x-axis contains the six input polarisation states of the imaging photon, while each row corresponds to the sample acting as different polarisers: horizontal, vertical, $+45^\circ$, and $-45^\circ$ linear polarisers, and polarisation-independent filters with $0\%$, $50\%$, and $100\%$ transmittance. The principal transmittances $T_f$ and $T_s$, and the reference axis orientation $\theta_{\mathrm{axis}}$, are indicated in each panel on the top right. All samples are modelled with zero retardance.}

    \label{fig.det-abs}
\end{figure}

\begin{figure}
    \centering
    \includegraphics[width=0.9\linewidth]{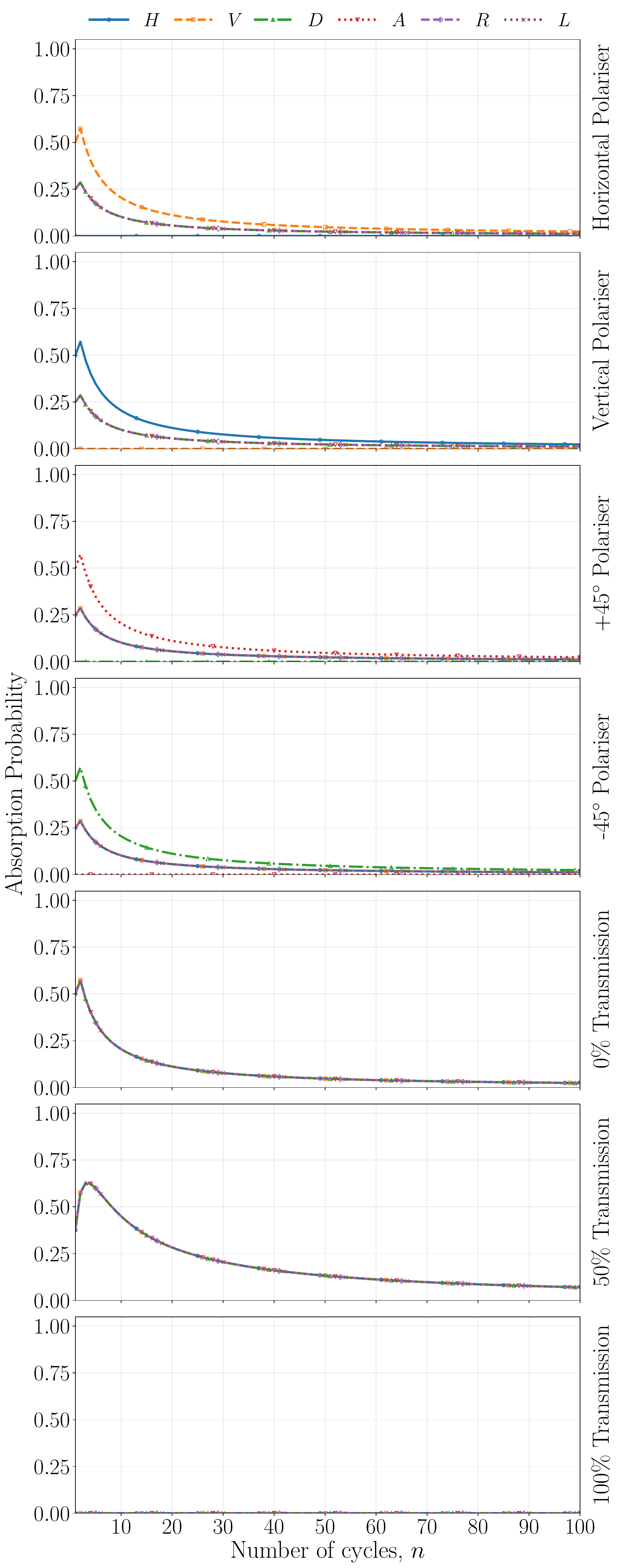}
    \caption{Absorption probabilities for the samples and input polarisation states considered in Fig.~\ref{fig.det-abs}, as a function of the total number of interferometer cycles $n$. The different lines correspond to the six input polarisation states $H$, $V$, $D$, $A$, $R$ and $L$.} 

    \label{fig.abs}
\end{figure}

\begin{figure}
    \centering
    \includegraphics[width=0.90\linewidth]{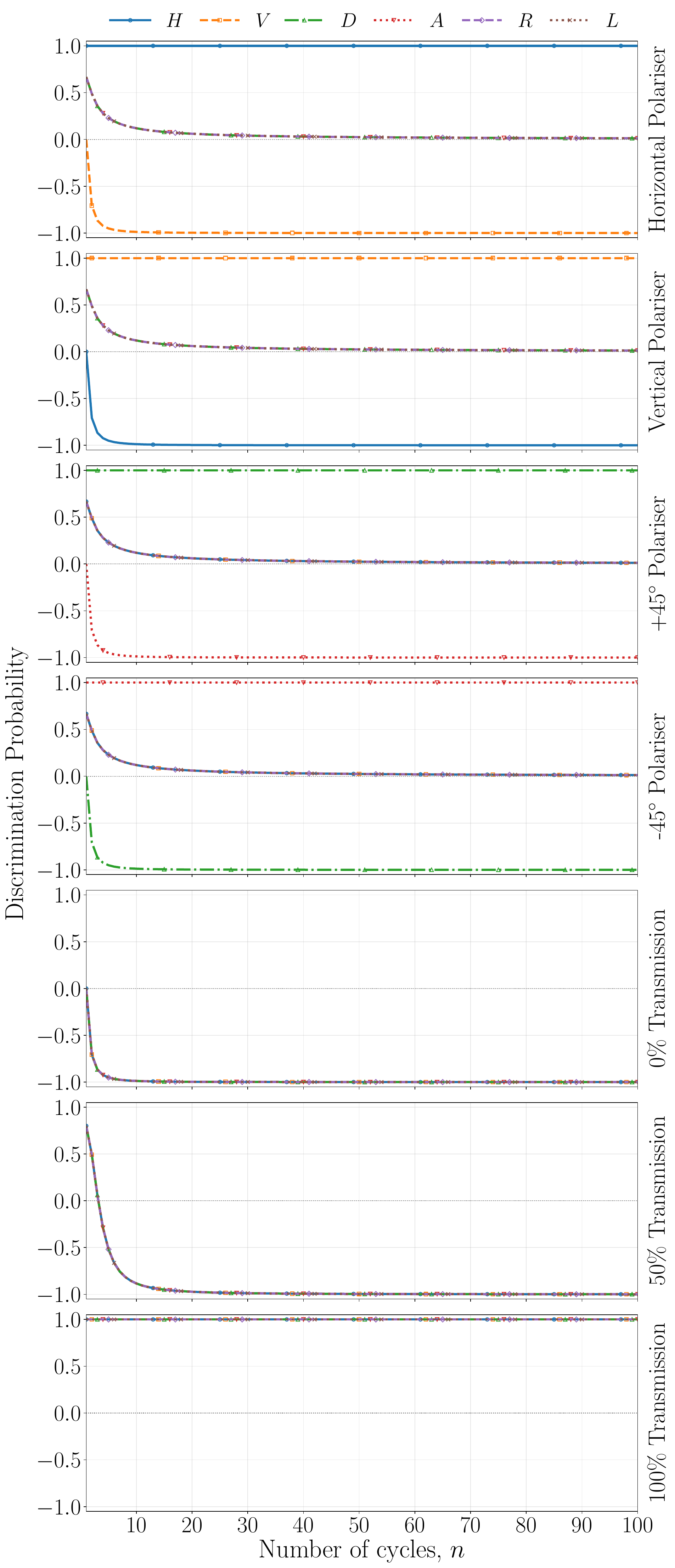}
    \caption{Port asymmetry $\mathcal{D}$ (labelled discrimination in the figure), defined as the normalised difference between outputs at the two detectors, $[P(\mathrm{ICCD0})-P(\mathrm{ICCD1})]/[P(\mathrm{ICCD0})+P(\mathrm{ICCD1})]$, for the samples and input polarisation states considered in Fig.~\ref{fig.det-abs}, as a function of the total number of interferometer cycles $n$. The different lines correspond to the six input polarisation states $H$, $V$, $D$, $A$, $R$ and $L$.} 

    \label{fig.dis}
\end{figure}

\subsubsection{Results}\label{sec:results}
The present study is restricted to linear diattenuators with zero retardance, characterised by their principal transmittances \(T_f\) and \(T_s\) respectively, along the principal axes $f$ and $s$, and by the orientation $\theta_{\mathrm{axis}}$ of the $f$ axis relative to the horizontal. Mueller matrix reconstruction is performed within this assumed sample class, and Sec.~\mbox{\ref{sec:beyond}} shows how the same mechanism responds to retardance and depolarisation. If a horizontally polarised imaging photon is incident on a sample which acts as an ideal horizontal polariser, the sample will transmit the incident photons without attenuation. This is true in general whenever the input polarisation is aligned with the fully transmitting axis of the sample. For these cases, the protocol behaves as in the no-sample case (see Appendix~\mbox{\ref{App:NoSample}}) and the photon exits through ICCD0 with unit probability; this is also already reflected in the $100\%$ transmitting sample in Fig.~\ref{fig.det-abs}. This behaviour, together with the detection and absorption probabilities for each of the six input polarisation states interacting with the different samples, is shown in Fig.~\ref{fig.det-abs}. This step-like behaviour is a feature of the Zeno limit: for large $n$ the output port depends almost entirely on whether a polarisation component is lossless or lossy. The amount of loss is encoded mainly in the small absorbed fraction (for $n=100$ the ICCD1 probability is $0.93$, $0.97$ and $0.98$ for $T=0.5$, $0.1$ and $0$). Resolving partial diattenuation therefore favours moderate $n$; the reconstruction below is demonstrated for $T\in\{0,\tfrac12,1\}$ without noise, and its sensitivity to intermediate $T$ under shot noise is quantified by the Fisher-information analysis of Sec.~\mbox{\ref{sec:fisher}}.

We calculate the detection probability at each detector from the corresponding Stokes parameter $S_0$, which is equivalent to the trace of the output density matrix associated with that detector. Absorption is calculated as $P_{\mathrm{abs}}=1-P(\mathrm{ICCD0})-P(\mathrm{ICCD1})$. The probabilities are not renormalised at detection by total surviving photon fraction to account for absorption. The results in Fig.~\ref{fig.det-abs} assume ideal optical components and exclude detector inefficiency and parasitic losses. In Fig.~\ref{fig.abs}, we show the fraction of incident photons absorbed by the sample as a function of the number of interferometer cycles. As discussed before, we can see when input polarisation of incident photons is aligned with the transmission axis of the sample, no absorption takes place as the sample transmits the photons completely in this ideal case.  For the polarisers, absorption is largest for the input orthogonal to the transmitting axis, peaking at $n=2$ and then decreasing as $n$ increases (the $50\%$ filter peaks at $n=3$). The remaining input states exhibit identical decreases, at half that value, in absorption as $n$ increases. This is because equal amounts of input polarisation lie along the  transmitting and absorbing axes of the sample. The simulation is consistent with absorption by the sample being suppressed due to quantum Zeno effect under ideal optical conditions, with zero sample retardance, and ignoring any other noise that could also account for some of the absorption. In Fig.~\ref{fig.dis}, we show that port asymmetry,
\begin{equation}
    \mathcal{D}= \frac{P(\mathrm{ICCD0})-P(\mathrm{ICCD1})}{P(\mathrm{ICCD0})+P(\mathrm{ICCD1})},
\end{equation}
is calculated as the difference between the intensities (detection probability per incident photon) at the two ICCDs and normalised by their sum.

\begin{figure}
    \centering
    \includegraphics[width=1\linewidth]{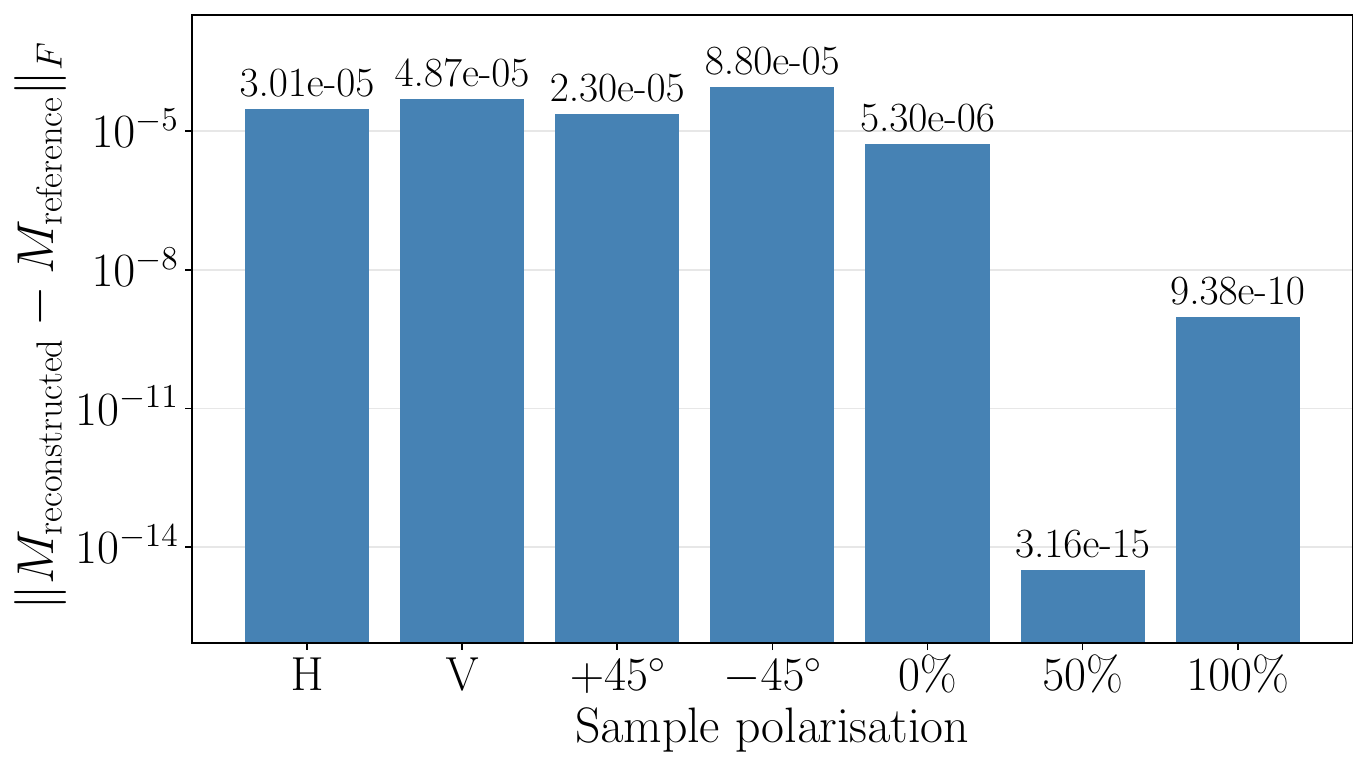}
    \caption{Frobenius norm of the difference between the reconstructed and reference sample Mueller matrices, $E_F=\lVert M_{\mathrm{rec}}-M_{\mathrm{ref}}\rVert_F$, for the seven samples of Fig.~\ref{fig.det-abs}, from the fitted sample parameters (noiseless simulation; the values reflect the optimiser tolerance). The y-axis uses a logarithmic scale. }

    \label{fig.error}
\end{figure}

Following the reconstruction procedure described in Sec.~\ref{Reconstructing Mueller matrix}, we use the best-fit sample parameters to obtain the Mueller matrices. We compare the reconstructed and reference Mueller matrices using the Frobenius norm of their difference, $E_F=\|M_{\mathrm{rec}}-M_{\mathrm{ref}}\|_F$ and report this error in Fig.~\ref{fig.error}. As the simulations are noiseless, $E_F$ quantifies numerical reconstruction accuracy rather than statistical uncertainty.

\section{Discussion}

\subsection{Modelling Detection/Interaction Rates}\label{sec:disc}

\begin{figure}[ht]
    \centering
    \includegraphics[width=0.95\linewidth]{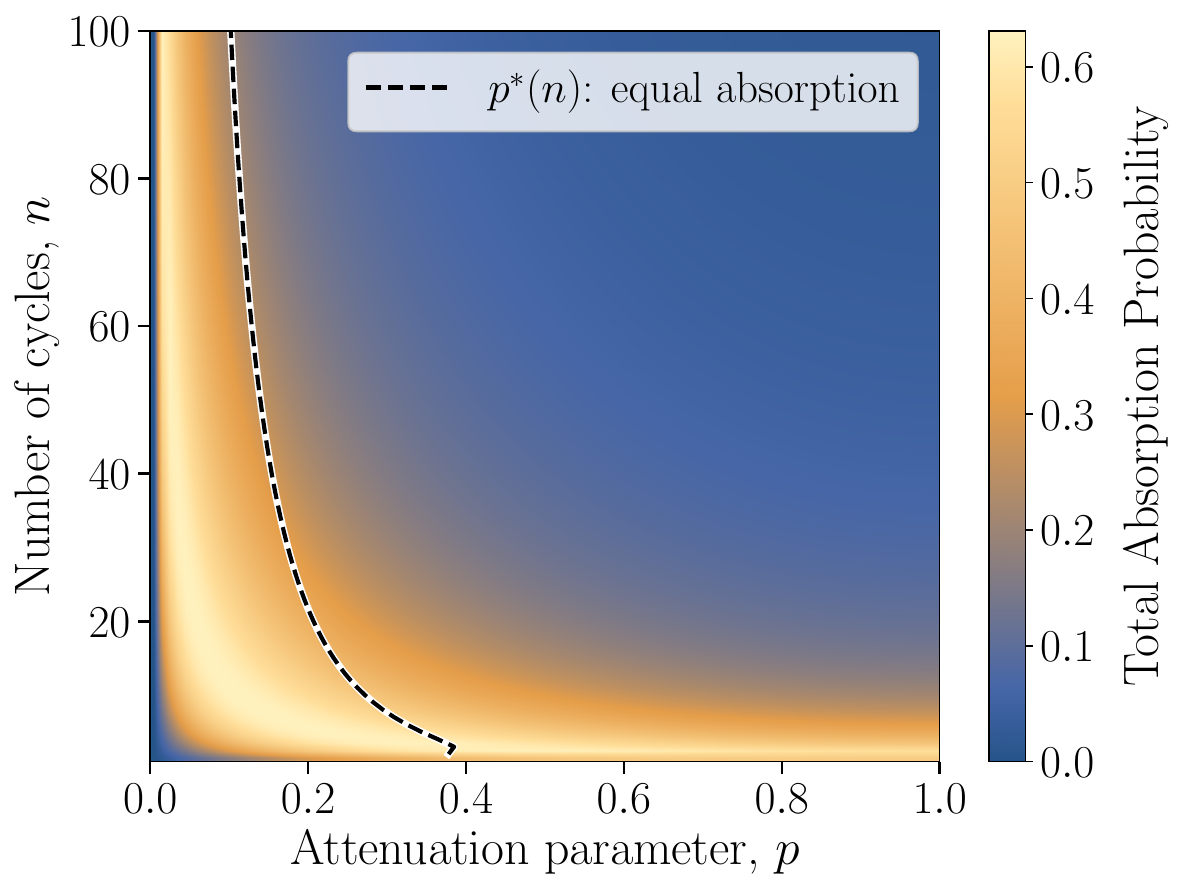}
    \caption{Probability of the sample absorbing the probe photon in an overall run of the protocol, as a function of $p$, the attenuation parameter of that polarisation-pixel mode (amplitude attenuation $1-p$ per pass, so single-pass intensity transmittance $(1-p)^2$ and single-pass absorption $1-(1-p)^2$), and $n$, the number of interferometer cycles in the run. The dashed curve $p^*(n)$ marks where the protocol’s absorption equals the single-pass absorption, \(P_{\mathrm{abs}}^{\mathrm{protocol}}=1-(1-p)^2\), with $p^*\approx0.102$ at $n=100$. To the right of this curve the protocol absorbs less than a single pass; to the left it absorbs more.} 
    \label{fig:TotAbsorbed}
\end{figure}

\begin{figure}[ht]
    \centering
    \includegraphics[width=0.95\linewidth]{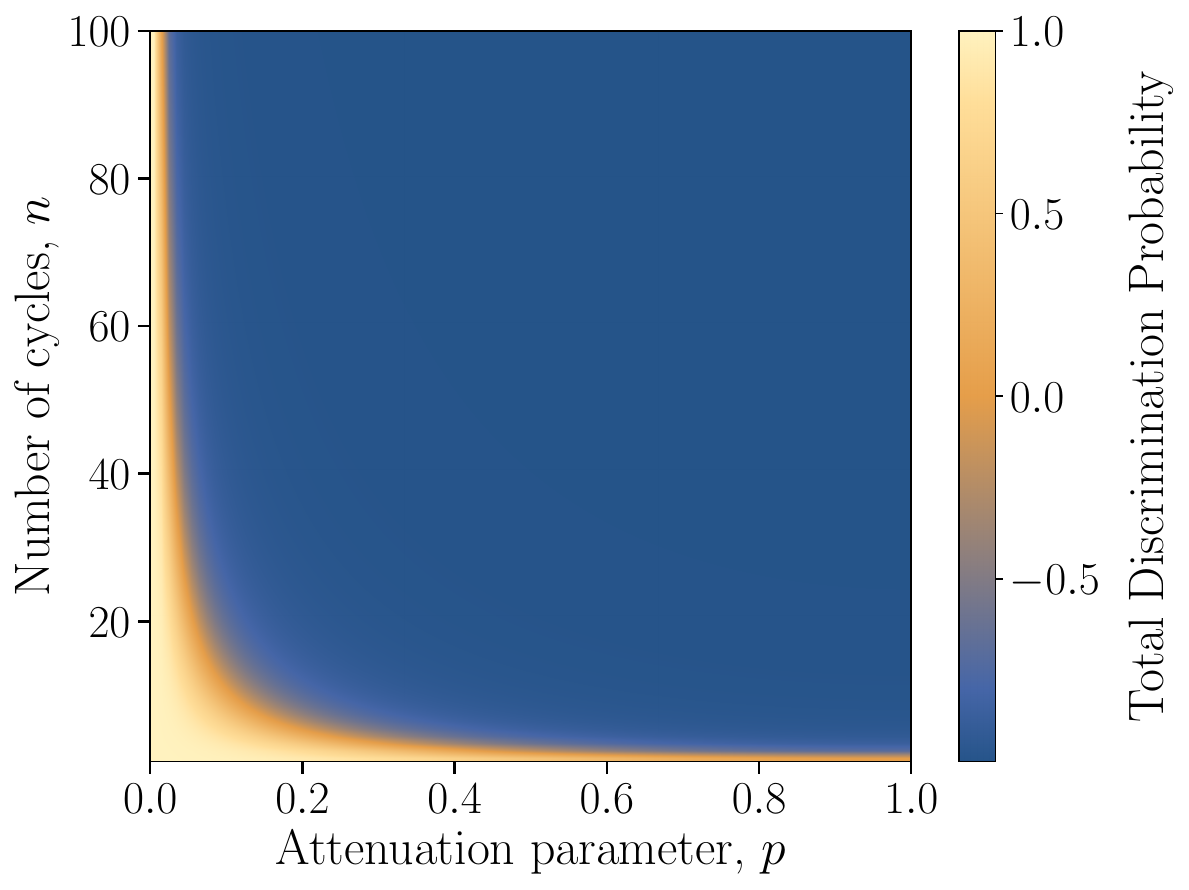}
    \caption{Port asymmetry $\big[P(\text{ICCD0})-P(\text{ICCD1})]/[P(\text{ICCD0})+P(\text{ICCD1})\big]$ (labelled discrimination in the figure), where ICCD1 is the port reached when the mode is blocked and ICCD0 the port reached when it is transmitted, so the no-sample case $p=0$ gives $1$, as a function of $p$, attenuation parameter (amplitude attenuation $1-p$ per pass and single-pass absorption $1-(1-p)^2$), and $n$, the number of interferometer cycles in the run.}
    \label{fig:Discrim}
\end{figure}

To evaluate the efficacy of the protocol, we need a way to describe the probability of the photon either arriving in one of the detectors, or being absorbed by the sample.

To do this, we consider one polarisation component of one pixel, which we call a polarisation-pixel mode. We model the diattenuation of this polarisation mode for a single pass through the object as an amplitude attenuation by a factor $1-p$, i.e. the Jones factor $\mathrm{diag}(1,\,1-p)$ on the (empty, sample) path modes, so that the single-pass intensity transmittance is $T=(1-p)^2$, the single-pass absorption probability is $1-(1-p)^2$, and $p=1$ is an opaque mode. This gives us the probability of the object absorbing the photon (Fig.~\ref{fig:TotAbsorbed}), and the probability of the photon arriving in ICCD0 minus the probability of it arriving in ICCD1 (the port reached when the mode is blocked) (Fig.~\ref{fig:Discrim}).

Fig.~\mbox{\ref{fig:TotAbsorbed}} shows that, as $n$ increases, the probability of the photon being absorbed by the sample falls below the single-pass absorption of a standard polarimetry protocol, $1-(1-p)^2$, but only for strongly attenuating modes. At $n=100$ this holds for $p\gtrsim 0.1$, i.e. per-pass intensity transmittance below about $0.8$ (the crossover transmittance $T^*$ is $0.53$, $0.81$ and $0.93$ at $n=10$, $100$ and $1000$). For weakly attenuating modes the repeated passes make the protocol absorb more than a single pass, up to a peak absorption of about $0.63$ on the ridge at $p\approx 1/n$. A lossless mode is traversed on every cycle. The suppression of absorption is therefore a property of strongly absorbing polarisation components, and the samples that benefit are those containing such components, for instance the polarisation-sensitive metasurfaces of Sec.~\mbox{\ref{sec:metasurf}}. This is the polarisation analogue of the known result for semi-transparent objects, that absorption-free discrimination needs one hypothesis to be full transparency \mbox{\cite{Mitchison2001Semitransparent,Thomas2014Semitransparency}}.

\begin{figure}
    \centering
    \includegraphics[width=\linewidth]{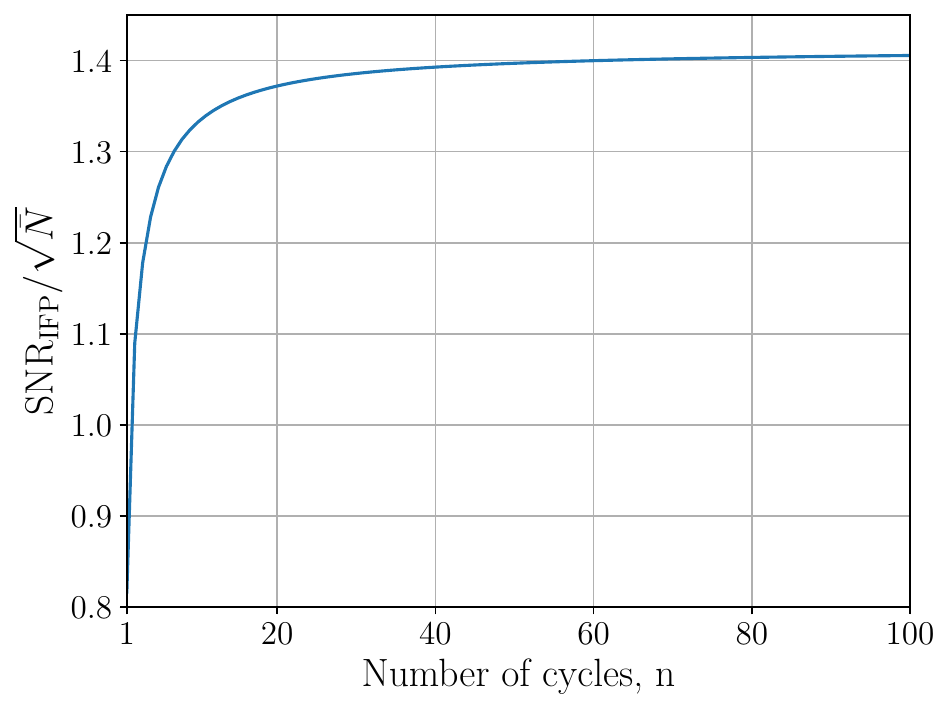}
    \caption{Signal-to-noise ($\mathrm{SNR_{IFP}}$) of the protocol for each polarisation-pixel mode as a function of $n$, the number of interferometer cycles, normalised by dividing by $\sqrt{\bar{N}}$, the square root of the average number of photons emitted into that mode. With this normalisation the SNR of standard polarimetry is 1, so values above 1 indicate that the protocol outperforms standard polarimetry per emitted photon. The curve saturates at $\sqrt{2}$, the gain from using both output ports.}
    \label{fig:SNR}
\end{figure}

\subsection{Signal-to-Noise Ratio}

The signal-to-noise ratio (SNR) \cite{OSullivan2010SNR,Brida2011SNR,Genovese2016ApplicationsQImaging}, a useful measure of the efficacy of an imaging system, is given by
\begin{equation}
    \text{SNR}=\frac{\vert\Delta\bar{I}\vert}{\sigma(\Delta\bar{I})},
\end{equation}
where $\Delta\bar{I}$ is the maximal difference in average intensity values observed by a detector, and $\sigma(\Delta\bar{I})$ is the standard deviation in this difference \mbox{\cite{Brida2011SNR}}.

For standard polarimetry, an average of $\bar{N}$ photons (those generated in a given time interval by an SPDC source) interrogate an object. This spontaneous parametric down-conversion has thermal statistics for rate of emission (which look Poissonian averaged over many temporal modes). Therefore, considering each polarisation-pixel mode, if we take $\Delta I$ as the difference in number of photons arriving at our detector between when the object is maximally transmissive to that polarisation in that pixel, and when it is maximally absorptive to that polarisation in that pixel; this difference is $\bar{N}$, and the SNR is
\begin{equation}
\begin{split}
\text{SNR}_{\text{Pol}}=\frac{\vert\Delta\bar{I}\vert}{\sigma(\Delta\bar{I})} = \frac{\bar{N}}{\sqrt{\bar{N}}} = \sqrt{\bar{N}}.
    \end{split}
\end{equation}

\begin{figure*}
    \centering
\includegraphics[width=0.9\linewidth]{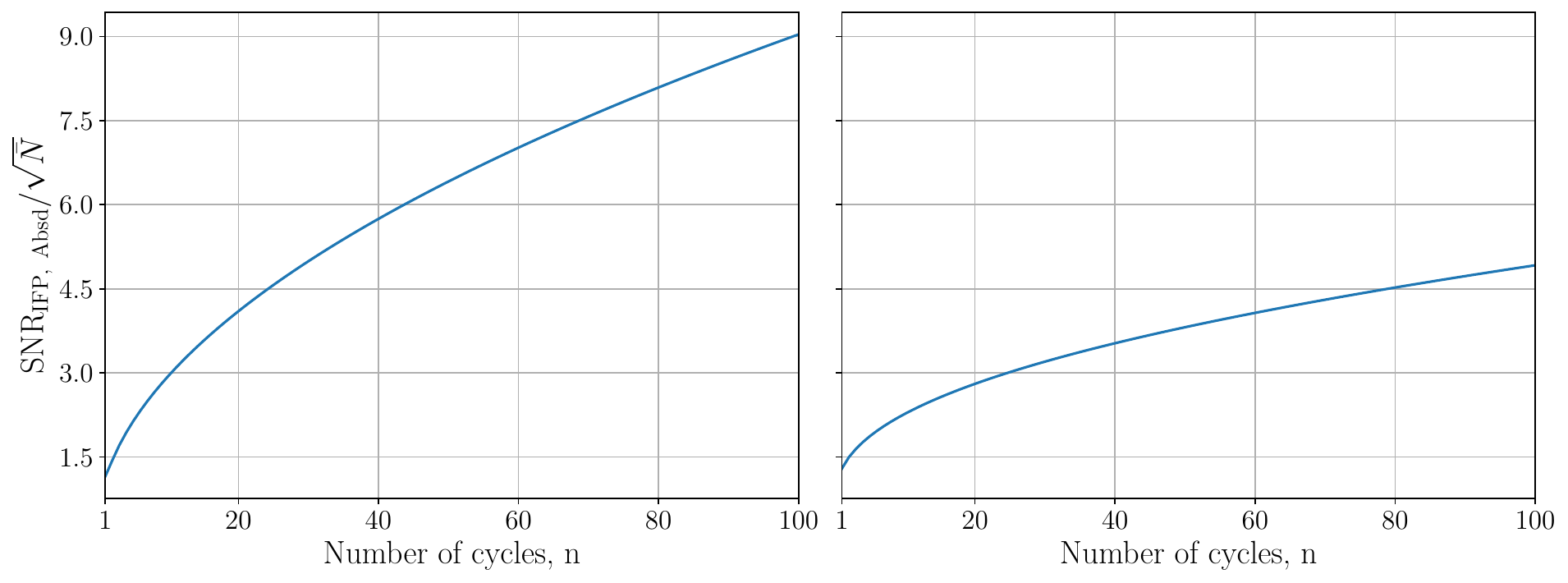}
    \caption{SNR per absorbed photon, $\mathrm{SNR}_{\mathrm{IFP, \ Absd}}=g(n)\sqrt{\bar N}$, for each polarisation-pixel mode as a function of $n$, the number of interferometer cycles. The plot shows $g(n)$, that is $\mathrm{SNR}_{\mathrm{Absd}}$ divided by $\sqrt{\bar{N}}$, the square root of the average number of photons emitted into that mode, with the absorbed fraction evaluated for an opaque sample at each cycle count (left), and with the absorbed fraction averaged over attenuation parameter $p$ (right) as described in the text. In these units, for detection of a fully opaque object, the single-pass measurement gives $\mathrm{SNR}/\sqrt{N_{\rm abs}}=1$, so values above 1 indicate a higher SNR than single-pass transmission at the same expected number of absorbed photons. For the alternative normalisation based on absorption averaged uniformly over $p\in[0,1]$, the single-pass measurement, which absorbs on average $2/3$ of its photons over the same range of $p$, has the value $\sqrt{3/2}\approx1.22$, so values above $1.22$ indicate that the protocol outperforms it per absorbed photon.}
    \label{fig:SNRPerAbsd}
\end{figure*}

For our above model, we define $\Delta N_{\text{ICCD1}}$ ($\Delta N_{\text{ICCD0}}$), $\Delta N_{\mathrm{ICCD}k}=N_{\mathrm{ICCD}k}^{\mathrm{transmissive}}-N_{\mathrm{ICCD}k}^{\mathrm{absorptive}}$ for $k=0,1$,  as the difference in photon numbers received at ICCD1 (ICCD0) between the object being transmissive to that polarisation in that pixel, and maximally absorptive to that polarisation in that pixel (which in each case is $\bar{N}$, the average number of photons emitted into that mode, times the difference in probability of a photon reaching that detector in each of those two cases). $\Delta N_{\text{ICCD1}}$ and $\Delta N_{\text{ICCD0}}$ will have opposite signs with $\Delta N_{\text{ICCD1}} < 0$ and $\Delta N_{\text{ICCD0}} > 0$. Therefore, the SNR for interaction-free polarimetry, $\text{SNR}_\text{IFP}$ can be written as
\begin{equation}
\begin{split}
    \text{SNR}_{\text{IFP}}&=\frac{\vert\Delta N_{\text{ICCD1}}-\Delta N_{\text{ICCD0}}\vert}{\sigma(\Delta N_{\text{ICCD1}}-\Delta N_{\text{ICCD0}})}\\
    &=f(n)\sqrt{\bar{N}} = f(n)\cdot\text{SNR}_{\text{Pol}},
    \end{split}
\end{equation}
where we plot $f(n)$ in \Cref{fig:SNR}; with Poissonian counts at the two detectors, $f(n)=\vert\Delta P_{\text{ICCD1}}-\Delta P_{\text{ICCD0}}\vert/\sqrt{\Sigma P_{\text{ICCD1}}+\Sigma P_{\text{ICCD0}}}$, where $\Sigma P_{Dk}$ is the sum of that detector's probabilities in the two cases, and $f(n)\to\sqrt{2}$ as $n\to\infty$ because the two ports then give perfectly anticorrelated signals.

However, more relevant to us is the SNR per absorbed photon. Averaging over the number of photons interacting with the sample as a function of $p$ and $n$ from \Cref{fig:TotAbsorbed}, or 100 equally-spaced values of $p$ in $[0,1]$, which gives $P_{\text{Absd}}(n)$, and dividing the SNR by $\sqrt{P_{\text{Absd}}}$, we get the SNR per absorbed photon
\begin{equation}
\begin{split}
    \text{SNR}&_{\text{IFP},\ \text{Absd}}=f(n)\sqrt{\bar{N}/P_{\text{Absd}}}\\
    &= g(n)\sqrt{\bar{N}}= g(n)\cdot\text{SNR}_{\text{Pol}},
\end{split}
\end{equation}

where we plot $g(n)$ in Fig.~\ref{fig:SNRPerAbsd}. For standard single-pass polarimetry, a fully transmissive sample transmits $N$ photons on average, whereas a fully opaque sample absorbs all $N$ incident photons. Therefore the difference in signal is $N$, with Poisson counting noise $\sqrt{N}$, giving $\mathrm{SNR}=\sqrt{N}$. With the expected number of absorbed photons for an opaque sample being $N_{\mathrm{Absd}}=N$, the single-pass SNR per absorbed photon is $\mathrm{SNR}/\sqrt{N_{\mathrm{Absd}}}=1$. At $n=100$, with $f(100)\approx1.41$ and corresponding $P_{\mathrm{Absd}}\approx0.024$ for an opaque sample, we get $g_{(100)}\approx9.04$ (see Fig.~\ref{fig:SNRPerAbsd} (left)), corresponding to approximately nine times the single-pass SNR at the same expected number of absorbed photons. Equivalently, at a fixed number $N_a=\bar N P_{\text{Absd}}$ of absorbed photons the SNR is $g(n)\sqrt{N_a}$. The comparison with standard polarimetry has to use the same average: a single pass absorbs $1-(1-p)^2$, which averages to $2/3$, so the standard SNR per absorbed photon is $\sqrt{3N_a/2}$ and the protocol's advantage at $n=100$ is $g(100)\sqrt{2/3}\approx 4.0$ rather than $g(100)\approx 4.9$ (see Fig.~\ref{fig:SNRPerAbsd} (right)). The advantage lies in discriminating an opaque from a transparent mode, not in estimating an intermediate transmittance. For that task a single-pass interaction-free measurement is known to give no gain per absorbed photon, and a multi-pass scheme gains nothing either, and generally loses, once each absorption is registered, as it is here through the herald \mbox{\cite{Massar2001MinimalAbsorption,Thomas2014Semitransparency,Wildfeuer2026Counterfactual,Hance2026Quasiprob}}; Sec.~\mbox{\ref{sec:fisher}} quantifies this for the polarimetric parameters.

\subsection{Fisher Information}\label{sec:fisher}
The SNR here just shows the ability of the protocol to distinguish a mode that is maximally transmissive from one that is maximally absorptive. More useful is how well the protocol determines changes in $p$ for each mode, which the Fisher information quantifies. The Fisher information per incident photon for loss estimation is analysed in \mbox{\cite{Allen2019Thesis}} for a single pass (Sec.~3.3) and for multiple passes (Sec.~4.5); here we compute it for the polarimetric parameters themselves. We compute the classical Fisher matrix of $(T_f,T_s,\theta_{\mathrm{axis}})$ for the measurement of Sec.~\mbox{\ref{Reconstructing Mueller matrix}} (six heralded inputs, three analysis bases at each port) and compare it with standard single-pass polarimetry using the same inputs and analysers, per incident and per absorbed photon. The samples below have their axes at $\pi/8$ to the input basis, a generic orientation; when an input polarisation coincides with a sample axis some outcomes have zero probability and the Fisher information is discontinuous there. For estimating transmittance $T_s$ of the absorbing axis, the protocol gives no more Fisher information per absorbed photon than a single pass. At $n=100$, the protocol gives $0.13$ to $0.94$ times the single-pass value for $T_f=1$ and $T_s\le0.5$, falling to $0.04$ when the transmitting axis is lossy with $T_f=0.95$. Determining how much of a polarisation component is absorbed by a sample requires photons to pass through it as for semi-transparent objects in general \mbox{\cite{Mitchison2001Semitransparent,Thomas2014Semitransparency,Wildfeuer2026Counterfactual,Hance2026Quasiprob}}, and the protocol suppresses this absorption, which makes the detection outcomes less sensitive to changes in $T_s$.

\begin{figure}[htbp]
    \centering
    \includegraphics[width=\columnwidth]{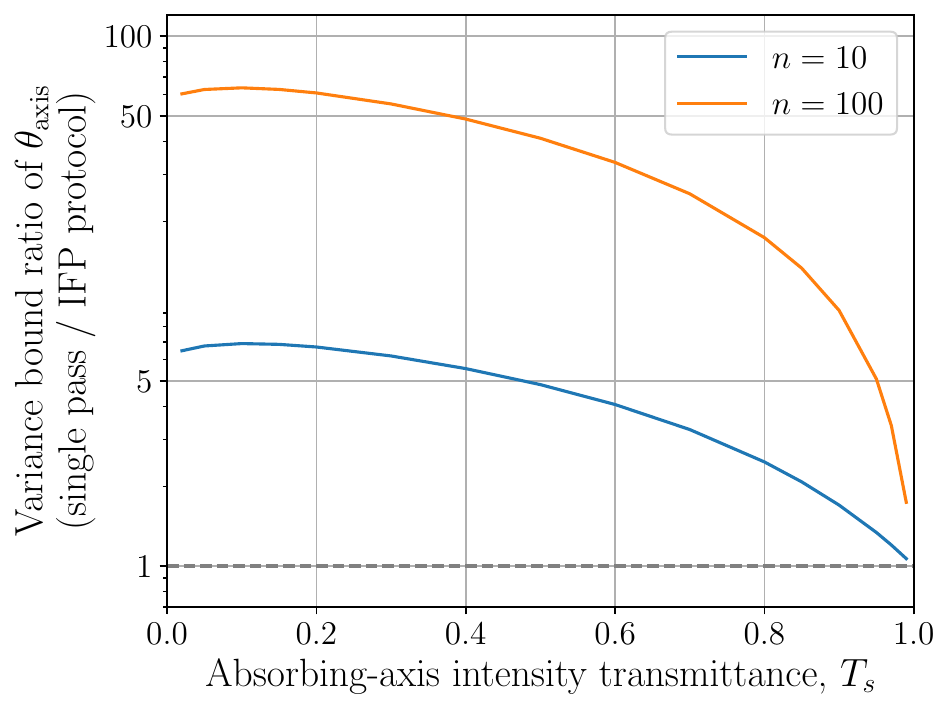}
    \caption{Ratio of the Cram\'er--Rao variance of the axis orientation $\theta_{\mathrm{axis}}$ per absorbed photon, single-pass polarimetry over the protocol, against the transmittance $T_s$ of the absorbing axis, for a sample with one lossless axis ($T_f=1$) at $\pi/8$ to the input basis, from the full Fisher matrix of $(T_f,T_s,\theta_{\mathrm{axis}})$ over the six inputs and three analysis bases of Sec.~\mbox{\ref{Reconstructing Mueller matrix}}, for $n=10$ and $n=100$. Above the dashed line (ratio $1$) the protocol locates the axis with fewer absorbed photons; the gain is largest for strongly absorbing axes and remains above unity up to $T_s=0.99$ at $n=100$.}
    \label{fig:fisheraxis}
\end{figure}

For the axis orientation, the information comes from which port the photon leaves, and here the protocol gains. We refer to the Cram\'er--Rao bound on the variance of $\theta_{\mathrm{axis}}$, given by the $\theta\theta$ element of the inverse of the Fisher matrix per incident photon $F$, as the Cram\'er--Rao variance, $\operatorname{Var}(\hat{\theta}_{\mathrm{axis}})\geq (F^{-1})_{\theta\theta}/N,$ with $N$ as the number of incident photons.
With one lossless axis ($T_f=1$), the Cram\'er--Rao variance per absorbed photon is $60$, $41$ and $9$ times smaller than for a single pass at $T_s=0.02$, $0.5$ and $0.9$, respectively ($n=100$; Fig.~\mbox{\ref{fig:fisheraxis}}). Per incident photon, the corresponding Cram\'er--Rao variances are smaller by factors of $1.6$, $5.7$ and $31$, respectively. We get this advantage because the single pass loses sensitivity to the axis orientation as the contrast $1-T_s$ vanishes while the Zeno interferometer still steers the lossy component to ICCD1. For an ideal polariser and a linearly polarised input, the port counts of both measurements obey Malus's law with a single pass Fisher information of $F_\theta=4$ per incident photon.
For the protocol, $F_\theta=3.98$ at $n=10$ and $F_\theta=4.00$ at $n=100$, approaching the single-pass value at large $n$. The protocol only absorbs $1-P_{\mathrm{surv}}(n)\approx\pi^2/4n$ of the single-pass fraction. We define the per-incident gain as the ratio of the Cram\'er--Rao variance bounds at equal incident-photon number, $G_{\mathrm{inc}}=(F_{\mathrm{single}}^{-1})_{\theta\theta}/{(F_{\mathrm{protocol}}^{-1})_{\theta\theta}}$. As transmittance $T_s\to0$, the per-incident gain that remains, $G_{\mathrm{inc}}\approx1.3$ at $n=100$, comes from analysing the polarisation of the photons at ICCD1, which the single pass absorbs. An adaptive single-pass strategy can reduce absorption by moving the input polarisation closer to the transmitting axis. Using the same polarisation would also reduce absorption in the protocol. However, the protocol absorbs only a fraction $\pi^2/4n$ of the photons that would be absorbed in a single-pass measurement, for the same number of incident photons. Thus, this adjustment that moves the input towards the transmitting axis does not remove the gain.

The gain requires the transmitting axis to be lossless to well within $1/n$ per pass: for $T_s=0.5$ at $n=100$ the $41$-fold gain per absorbed photon falls to $2.9$ at $T_f=0.99$ and to $1.1$ at $T_f=0.95$ ($1.2$ per incident photon), and at $T_f=0.8$ it is gone, since both axes are then too lossy for the Zeno sorting to separate them. The protocol is thus a polarimeter for samples with a lossless axis: it locates the absorbing axis of such a polariser-like feature with far fewer absorbed photons and distinguishes absorbing from transmitting components (Sec.~\mbox{\ref{sec:disc}}), but it does not measure partial transmittances more efficiently per absorbed photon than a single pass.

\subsection{Beyond Diattenuation}\label{sec:beyond}
The protocol sorts a polarisation component to ICCD1 when the amplitude leaking into the sample arm each cycle cannot, on its return, interfere constructively with the amplitude in the empty arm. Absorption is one way to spoil that interference, and a phase shift and decoherence are others, so retardance and depolarisation are detected by the same mechanism. We illustrate this in the model of Sec.~\mbox{\ref{sec:numerical}} with the sample described as a quantum operation on the path-1 polarisation; Table~\mbox{\ref{tab:beyond}} collects the numbers. For a lossless linear retarder, with the interferometer balanced on its fast axis, the slow component is sent to ICCD1 with probability $1/2$ at a retardance per pass of about $1.25/n$ and held almost completely well above it: at $n=100$ a retardance of $0.1$ sends it to ICCD1 with probability $0.99$ after $2.3$ passes through the sample, and a quarter-wave retardance per pass flips the sign of the leaked amplitude each cycle, so that the component never transfers and passes the sample $\pi^2/8n$ times; the output polarisation stays pure at both ports. A depolariser, modelled as the depolarising channel \mbox{\cite{nielsen_chuang_2010}} applied once per pass (one of the three Pauli operations on the sample-arm polarisation with total probability $q$, drawn independently of the photon's path, with Mueller matrix $\mathrm{diag}(1,1-4q/3,1-4q/3,1-4q/3)$), is sorted as well: a complete depolariser ($q=3/4$) reaches ICCD1 with probability $0.945$ at $n=100$, and the two ports differ, the photon at ICCD1 keeping its input polarisation (purity $0.999$) while the photon at ICCD0 is fully depolarised. However, decoherence does not remove the leaked amplitude, so the photon makes a random walk between the arms and passes a complete depolariser $5.6$ times, and a partial one ($q=0.3$) $13$ times, more often than in a single pass.

\begin{table}[t]
\caption{Retardance and depolarisation in the protocol (this subsection), with the inputs and analysis of Sec.~\mbox{\ref{Reconstructing Mueller matrix}}. The lossless retarder is probed on its slow axis with the interferometer balanced on the fast axis, the depolariser with an $H$ input; ``single'' is single-pass polarimetry with the same settings, $\delta$ is the retardance per pass, and the variance ratios are Cram\'er--Rao variances of the named parameter, per absorbed or per incident photon.}
\label{tab:beyond}
\footnotesize
\begin{ruledtabular}
\begin{tabular}{p{5.2cm}cc}
 & $n=10$ & $n=100$ \\
\hline
\multicolumn{3}{l}{%
  \fcolorbox{black}{gray!10}{%
    \textbf{Lossless retarder, }$\boldsymbol{\delta=0.1}$%
  }%
} \\
\quad $P(\mathrm{ICCD1})$ & $0.35$ & $0.99$ \\
\cline{2-3}
\quad Passes through sample & $8.2$ & $2.3$ \\
\hline
\multicolumn{3}{l}{%
  \fcolorbox{black}{gray!10}{%
    \shortstack[l]{%
      \textbf{Lossless retarder, }$\boldsymbol{\delta=\pi/2}$\\
      \textbf{(quarter wave)}%
    }%
  }%
} \\
\quad $P(\mathrm{ICCD1})$ & $1$ & $1$ \\
\cline{2-3}
\quad Passes through sample ($\pi^2/8n$) & $0.12$ & $0.012$ \\
\hline
\multicolumn{3}{l}{%
  \fcolorbox{black}{gray!10}{%
    \textbf{Depolariser, }$\boldsymbol{q=3/4}$%
  }%
} \\
\quad $P(\mathrm{ICCD1})$ & $0.65$ & $0.945$ \\
\cline{2-3}
\quad Passes through sample & $3.7$ & $5.6$ \\
\cline{2-3}
\quad Purity at $\frac{\mathrm{ICCD0}}{\mathrm{ICCD1}}$ & $\frac{0.50}{0.93}$ & $\frac{0.50}{0.999}$ \\[3pt]

\hline
\multicolumn{3}{l}{%
  \fcolorbox{black}{gray!10}{%
    \textbf{Depolariser, }$\boldsymbol{q=0.3}$%
  }%
} \\
\quad Passes through sample & $5.9$ & $13$ \\
\hline
\multicolumn{3}{l}{%
  \fcolorbox{black}{gray!10}{%
    \shortstack[l]{%
      \textbf{Depolariser, }$\boldsymbol{q=0.1:}$\\
       $\boldsymbol{F_q,}$\ \textbf{protocol / single}%
    }%
  }%
} \\
\quad Per incident photon & $9.0$ & $7.6$ \\
\cline{2-3}
\quad Per pass through sample & $1.26$ & $0.31$ \\
\hline
\multicolumn{3}{l}{%
  \fcolorbox{black}{gray!10}{%
    \shortstack[l]{%
      \textbf{Diattenuating retarder,}\\
      $\boldsymbol{
        T_f=1,\ T_s=0.5,\ \theta_{\mathrm{axis}}=\pi/8
      }$
    }%
  }%
} \\
\quad Mean absorption, $\delta=0$\newline
\makebox[\linewidth][c]{(single pass $0.25$)} & $0.23$ & $0.035$ \\
\cline{2-3}
\quad Mean absorption, $\delta=\pi/4$ & $0.07$ & $0.007$ \\
\cline{2-3}
\quad Mean absorption, $\delta=\pi/2$ & $0.04$ & $0.004$ \\
\hline
\quad $\theta_{\mathrm{axis}}$ variance ratio,\newline
\makebox[\linewidth][c]{single / protocol ($\delta=0.6$)} & & \\
\quad Per absorbed photon & $11$ & $106$ \\
\cline{2-3}
\quad Per incident photon & $3.9$ & $4.3$ \\
\hline
\quad$\delta$ variance,
 protocol / single & & \\
\quad Per absorbed photon & $1.8$ & $15$ \\
\cline{2-3}
\hline
\multicolumn{3}{l}{%
  \fcolorbox{black}{gray!10}{%
    \shortstack[l]{%
      \textbf{Diattenuating retarder,}\\
      $\boldsymbol{T_f=0.95,\ T_s=0.5,}$\\
      $\boldsymbol{\theta_{\mathrm{axis}}=\pi/8,\ \delta=0.6}$
    }%
  }%
} \\

\quad$\delta$ variance,
protocol / single & & \\
\quad Per absorbed photon & $4.4$ & $260$ \\
\cline{2-3}
\quad Per incident photon &  $4.4$ & $270$ \\
\hline
\quad Fisher-matrix condition number & $10^{3.4}$ & $10^{5.4}$ \\
\hline
\multicolumn{3}{l}{%
  \fcolorbox{black}{gray!10}{%
    \textbf{Fit sample }$\boldsymbol{(0.9,\,0.4,\,\pi/5,\,0.6)}$%
  }%
} \\
\quad Fisher-matrix condition number & $10^{3.1}$ & $10^{4.6}$ \\
\end{tabular}
\end{ruledtabular}
\end{table}

A lossless retarder or depolariser absorbs no energy, so the motivation of this paper does not apply to them on their own (interaction-free polarimetry of a transparent birefringent sample is the case that Ref.~\mbox{\cite{Hance2026Quasiprob}}, Sec.~VII, leaves outside its absorption bounds; the lossless retarder above is a first step towards it). The relevant case is a sample that both diattenuates and retards, a linear diattenuating retarder in the terminology of Lu and Chipman \mbox{\cite{Lu1996PolarDecomposition}}. There, a retardance on the absorbing axis adds to the Zeno inhibition and lowers the absorption further, from $0.035$ to $0.007$ for $T_f=1$, $T_s=0.5$ and a retardance of $\pi/4$ at $n=100$ (mean over the six inputs, which is independent of the axis orientation), and the axis orientation keeps the advantage of Sec.~\mbox{\ref{sec:fisher}}, with a Cram\'er--Rao variance per absorbed photon $106$ times below a single pass at $n=100$ for $T_f=1$, $T_s=0.5$, a retardance of $0.6$ and the axis at $\pi/8$ as in Sec.~\mbox{\ref{sec:fisher}}. The retardance itself is detected but poorly quantified: for a sample with both axes lossy ($T_f=0.95$, $T_s=0.5$, retardance $0.6$) its Cram\'er--Rao variance per absorbed photon is $4.4$ ($n=10$) to $260$ ($n=100$) times that of a single pass ($1.8$ and $15$ times for $T_f=1$), because the photon is kept away from the axis that would reveal it. The depolarisation strength is learnt faster per incident photon ($9$ times at $q=0.1$, $n=10$) but, per pass through the sample, at most $1.3$ times faster ($n=10$) and more slowly than in a single pass at $n=100$. The port-resolved Stokes data separate the three effects, since loss changes the trace of the port operations, retardance rotates a still-pure output and depolarisation lowers the purity at ICCD0 only, and the quantum-operation description of Sec.~\mbox{\ref{sec:quantumrep}} covers all three. The four parameters $(T_f,T_s,\theta_{\mathrm{axis}},\delta)$ of a linear diattenuating retarder are then identifiable from the port-resolved data (nonsingular Fisher matrix for the samples above, and a unique zero-residual fit within the parameter box for a test sample at $(0.9,\,0.4,\,\pi/5,\,0.6)$), with a conditioning that worsens with $n$ (condition number $10^{3.4}$ at $n=10$ and $10^{5.4}$ at $n=100$ for the $T_f=0.95$ sample), the same large-$n$ insensitivity as for partial transmittance. Quantitative work on such samples thus favours moderate $n$, while the location of absorbing or retarding axes benefits from large $n$.

\section{Conclusion}
In this paper, we gave an extension of counterfactual imaging to imaging of polarising objects, and then to the reconstruction of the Mueller matrix of a linear diattenuator with zero retardance. For strongly absorbing polarisation components (per-pass intensity transmittance below about $0.8$ at $n=100$) the protocol absorbs less energy than single-pass polarimetry, by a factor of order $n$ for an opaque component (from unity to $\pi^2/4n$). This is the benefit sought for delicate samples probed with high-frequency radiation. For weakly absorbing components it absorbs more, so the method suits strongly absorbing polarisation features. In our model the SNR per emitted photon is at most $\sqrt2$ times the standard value (Fig.~\mbox{\ref{fig:SNR}}), and the SNR per absorbed photon grows with $n$ (Fig.~\mbox{\ref{fig:SNRPerAbsd}}). The Fisher-information analysis of Sec.~\mbox{\ref{sec:fisher}} shows where the advantage lies: the protocol locates the absorbing axis of a sample whose transmitting axis is lossless with far fewer absorbed photons than a single pass, while partial transmittances are measured no more efficiently per absorbed photon than by a single pass. The same sorting mechanism responds to retardance and depolarisation (Sec.~\mbox{\ref{sec:beyond}}), so the protocol extends to diattenuating retarders with a lossless axis, locating the absorbing axis with few absorptions while leaving the quantitative retardance to a single-pass measurement.

A natural first experiment is the imaging of a polarisation-sensitive metasurface, which stores several polarisation-dependent images, with multi-pixel imaging as in \mbox{\cite{Hance2021CFGI,Palici2022IFMMultimode}}. Altuzarra et al.~imaged such a metasurface with polarisation-correlated and polarisation-entangled photon pairs, the image that appears being selected by the polarisation measured on the herald photon \mbox{\cite{Altuzarra2019Metasurf}}, a heralded protocol comparable to ours though not interaction-free. Repeating that experiment with our protocol would allow a direct comparison of the absorbed energy at equal image quality.\label{sec:metasurf}

\textit{Acknowledgements:}
We thank ZhaoQiZhi Han for useful discussions. SC is funded by the International Graduate School in Quantum Technologies (IGSQT). JRH acknowledges support from Hiroshima University's Phoenix Postdoctoral Fellowship for Research, the University of York's EPSRC DTP grant EP/R513386/1, a Royal Society Research Grant (RG/R1/251590), and from their EPSRC Quantum Technologies Career Acceleration Fellowship (UKRI1217). JRH and JR acknowledge support from the UK Quantum Communications Hub, funded by the EPSRC grants EP/M013472/1 and EP/T001011/1.

\textit{Data and Code Availability:} Code reproducing all computations in this paper is available at \cite{OurRawData}.

\bibliographystyle{unsrturl}
\bibliography{ref.bib}
\clearpage
\onecolumngrid
\appendix

\section{Mapping Qubit Density Matrices to Stokes Parameters, and Quantum Operations to Mueller Matrices}
\begin{figure*}
\centering
    \includegraphics[scale=0.5]{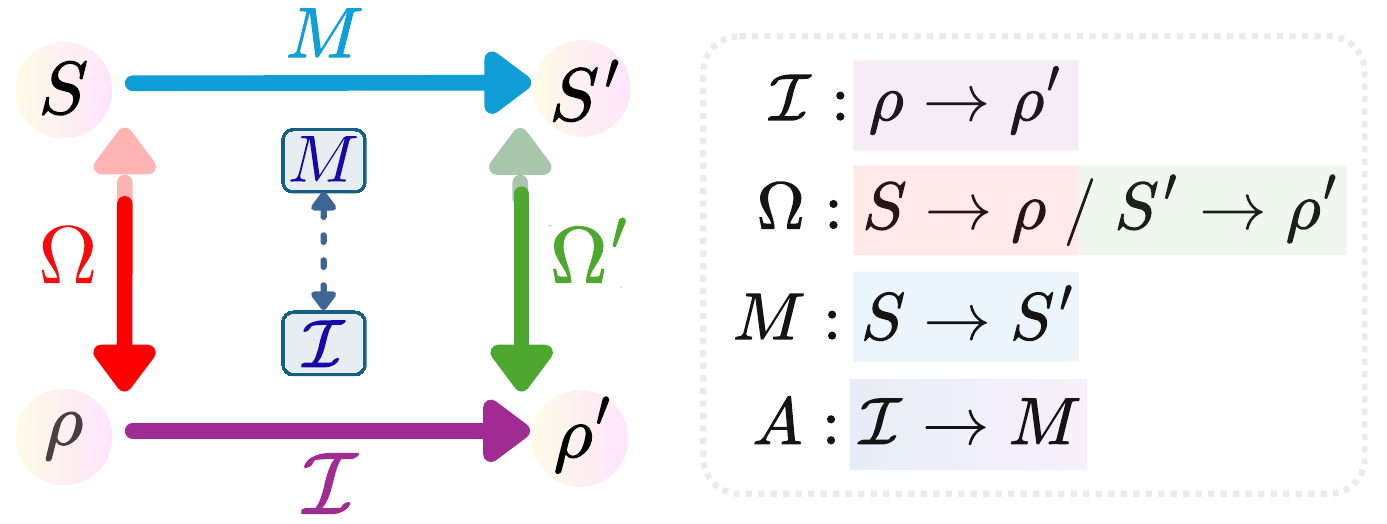}
    \caption{Commutative diagram of conversions $\Omega$ between Stokes vectors ($S$ and $S'$), and polarisation density matrices ($\rho$ and $\rho'$). The Mueller matrix $M$ represents the transformation between $S$ and $S'$, and the quantum operation $\mathcal{I}$ represents the transformation between $\rho$ and $\rho'$, so $M$ and $\mathcal{I}$ should implement equivalent transformations; the map $A$ takes $\mathcal{I}$ to $M$ through \mbox{Eq.~\eqref{eq:muller_qunatum}}.}
    \label{fig:map}
\vspace{-1.5em}
\end{figure*}

In this appendix, we describe the correspondence between classical Stokes vectors and quantum polarisation density matrices, and between classical Mueller matrices and quantum operations. This correspondence and its connection to Mueller matrices and quantum operations is illustrated by the commutative diagram in Fig.~\ref{fig:map}. The correspondence and the physicality conditions are standard \mbox{\cite{Aiello2007LinOptQMaps,Simon2010NonquantumEntanglement,Goldberg2020QPolarimetry}}; we give them here in the form used by the simulation, for the port-resolved, trace-non-increasing quantum operations of the interferometer. In Appendix~\ref{App:MapDenstoStokes}, we derive the transformation $\Omega$ from a Stokes vector to a polarisation density matrix, together with the inverse transformation $\Omega'$ from a density matrix to its corresponding Stokes vector. In Appendix~\ref{App:MuellerfromChannel}, we use this correspondence to derive the Mueller matrix associated with a given quantum operation $\mathcal{I}$. In Appendix~\ref{App:Deriving_Quantum Channel_Mueller}, we consider the inverse problem and reconstruct a quantum operation from a given Mueller matrix. Finally, in Appendix~\ref{Physical_val}, we establish the physicality conditions that the reconstructed operation must satisfy, including complete positivity and trace preservation or trace non-increase.

\subsection{Mapping Density Matrices to Stokes Parameters and vice versa}\label{App:MapDenstoStokes}

A density matrix describing a qubit can be
expanded in the Pauli basis as \cite{nielsen_chuang_2010}

\begin{equation}\label{eq:densdecomp}
    \rho = \frac{I + \vec{r} \cdot \vec{\sigma}}{2}
\end{equation}
where $\vec{r} = (r_x, r_y, r_z)$ are the components of the vector in the Bloch sphere, and  $\vec{\sigma}=(\sigma_x,\sigma_y,\sigma_z)$ represents the
three Pauli matrices.

After expanding the dot product in Eq.~\eqref{eq:densdecomp}, the density matrix can be written explicitly as
\begin{equation}
    \rho = \frac{1}{2} \left( I + r_x \sigma_x + r_y \sigma_y + r_z \sigma_z \right).
\end{equation}

The components of the Bloch vector, $r_x$, $r_y$, and $r_z$ are defined as $r_j = \mathrm{Tr}(\rho\, \sigma_j)$ for $j = x, y, z$. For a normalised polarisation state in the horizontal/vertical
basis, these correspond to the Stokes parameters in the order

\begin{equation}
    S_1 = r_z, \qquad S_2 = r_x, \qquad S_3 = r_y.
\end{equation}

For a normalised density matrix $\rho$, the zeroth Stokes parameter $S_0 = \mathrm{Tr}(\rho) = 1$.

Thus, the density matrix can also be written as

\begin{equation}\label{eq:rhoasStokes}
    \rho = \frac{1}{2} \left( S_0 I + S_1 \sigma_z + S_2 \sigma_x + S_3 \sigma_y \right).
\end{equation}

The decomposition in Eq.~\eqref{eq:rhoasStokes} provides us the basis for a transformation $\Omega$ from the Stokes vector for a single photon, to the density matrix for that photon and an inverse transformation $\Omega'$ from  density matrix to Stokes vector. To achieve the latter, we multiply the density matrix by the basis operator corresponding to each Stokes parameter and take the trace,
using $I$, $\sigma_z$, $\sigma_x$, and $\sigma_y$ for
$S_0$, $S_1$, $S_2$, and $S_3$, respectively.  As an explicit example, extracting the $S_0$ parameter from Eq.~\eqref{eq:rhoasStokes} is achieved by taking the trace of the density matrix with the identity operator. Because the Pauli matrices are traceless, $\operatorname{Tr}(\sigma_j) = 0$, and $\operatorname{Tr}(I^2) = 2$, the expansion simplifies to:
\begin{equation}
\begin{aligned}
    \operatorname{Tr}(\rho I) &= \frac{1}{2}\big( S_0\operatorname{Tr}(I^2) + S_1\operatorname{Tr}(\sigma_z) + S_2\operatorname{Tr}(\sigma_x) + S_3\operatorname{Tr}(\sigma_y) \big) \\
    &= S_0.
\end{aligned}
\end{equation}
Similarly, the remaining Stokes parameters are recovered by projecting the state onto the corresponding Pauli operators. For a given index $k \in \{0, 1, 2, 3\}$, associated with the basis matrices $(\sigma_0, \sigma_1, \sigma_2, \sigma_3) = (I, \sigma_z, \sigma_x, \sigma_y)$, we evaluate:
\begin{equation}
\begin{aligned}
    \operatorname{Tr}(\rho \sigma_k) &= \frac{1}{2}\big( S_0\operatorname{Tr}(\sigma_0\sigma_k) + S_1\operatorname{Tr}(\sigma_1\sigma_k) + S_2\operatorname{Tr}(\sigma_2\sigma_k) + S_3\operatorname{Tr}(\sigma_3\sigma_k) \big) \\
    &= S_k.
\end{aligned}
\label{stokesdensproof}
\end{equation}
This direct reduction to $S_k$ relies universally on the trace orthogonality condition $\operatorname{Tr}(\sigma_j\sigma_k) = 2\delta_{jk}$.

Therefore, we can obtain the Stokes parameters from the  density matrix as
\begin{equation}\label{eq:StokesfromDens}
S_0 = \mathrm{Tr}(\rho), \quad
S_1 = \mathrm{Tr}(\rho \sigma_z), \quad
S_2 = \mathrm{Tr}(\rho \sigma_x), \quad
S_3 = \mathrm{Tr}(\rho \sigma_y).
\end{equation}
with Pauli matrices $ \sigma_x = 
    \begin{pmatrix}
        0 & 1 \\
        1 & 0 \end{pmatrix}, \
    \sigma_y = 
    \begin{pmatrix}
        0 & -i \\
        i & 0
    \end{pmatrix}, \
    \sigma_z = 
    \begin{pmatrix}
        1 & 0 \\
        0 & -1
    \end{pmatrix}.
$

To verify our mapping, we can apply it to the general parametrisation of a normalised qubit density matrix:
\begin{equation}
    \rho_{\text{test}} = \begin{pmatrix} a & c \\ c^* & 1-a \end{pmatrix}, \quad 0 \leq a \leq 1, \quad |c|^2 \leq a(1-a).
\end{equation}

From Eq.~\eqref{eq:StokesfromDens}, we can calculate the Stokes parameters:
\begin{equation}
\begin{split}
S_0 &= \mathrm{Tr}(\rho) = 1 \\[0.5em]
S_1 &= \mathrm{Tr}(\rho \sigma_z)
= 2a-1 \\[1em]
S_2 &= \mathrm{Tr}(\rho \sigma_x)
=2\mathrm{Re}(c) \\[1em]
S_3 &= \mathrm{Tr}(\rho \sigma_y)
= -2\mathrm{Im}(c).
\end{split}
\end{equation}

Plugging these into the Pauli expansion from Eq.~\eqref{eq:rhoasStokes}:
\begin{equation}
\begin{aligned}
\rho &=
\frac{1}{2}
\left(
\begin{pmatrix} 1 & 0 \\ 0 & 1 \end{pmatrix}
+ (2a-1)
\begin{pmatrix} 1 & 0 \\ 0 & -1 \end{pmatrix}
+ 2\mathrm{Re}(c)
\begin{pmatrix} 0 & 1 \\ 1 & 0 \end{pmatrix}
- 2\mathrm{Im}(c)
\begin{pmatrix} 0 & -i \\ i & 0 \end{pmatrix}
\right) \\
&= \rho_{\text{test}}.
\end{aligned}
\end{equation}

This proves that the Pauli expansion reconstructs the original density matrix for any single qubit state.

\subsection{Derivation of Mueller Matrix from a quantum operation}
\label{App:MuellerfromChannel}
In this section we obtain the Mueller matrix from a quantum
operation acting on a density matrix. To allow for absorption by a sample and post selection, we allow this operation to be trace-non-increasing and retain the unnormalised output states at the ICCD detectors. Let us suppose a quantum operation $\mathcal{I}$ acting on density matrix $\rho$. By linearity, the outcome of this operation can be written as

\begin{equation}
\begin{aligned}
\rho'
&= \mathcal{I}(\rho)
 = \mathcal{I}\left(
    \frac{1}{2}\sum_{k=0}^{3}S_k\alpha_k
   \right)\\
&= \frac{1}{2}\sum_{k=0}^{3}
   S_k\,\mathcal{I}(\alpha_k).
\end{aligned}
\label{eq:rho_prime}
\end{equation}

where $\rho=\frac{1}{2}\sum_{k=0}^{3}S_k\alpha_k$ with 
$(\alpha_0,\alpha_1,\alpha_2,\alpha_3)
=(I,\sigma_z,\sigma_x,\sigma_y)$ as defined in \mbox{Eq.~\eqref{eq:rhoasStokes}}.

As established in Eq.~\ref{stokesdensproof}, we can compute the output Stokes parameters $S'_j$ using the trace and linearity:
\begin{equation}
\begin{aligned}
S'_j
&= \operatorname{Tr}(\rho'\alpha_j)\\
&= \frac{1}{2}\sum_{k=0}^{3}S_k\,
   \operatorname{Tr}\left[
   \alpha_j\mathcal{I}(\alpha_k)
   \right].
\end{aligned}
\label{eq:out_stokes_dense}
\end{equation}

This gives a linear transformation which is the explicit representation of the standard polarimetry equation $\mathbf{S}' = M\mathbf{S}$, 
given as 
\begin{equation}
S'_j = \sum_{k=0}^{3}M_{jk}S_k,
\end{equation}
where the Mueller matrix elements are defined as
\begin{equation}
M_{jk}
:= \frac{1}{2}\operatorname{Tr}\left[
\alpha_j\mathcal{I}(\alpha_k)
\right].
\label{eq:muller_qunatum}
\end{equation}

The action of the quantum operation $\mathcal{I}$ can be described using Kraus
operators $\{E_\ell\}$ as
\begin{equation}
\mathcal{I}(\alpha_k)
= \sum_\ell E_\ell\alpha_k E_\ell^\dagger.
\end{equation}

where for representing a passive and possibly lossy process, the operation must be completely positive and trace-non-increasing, satisfying the inequality $\sum_\ell E_\ell^\dagger E_\ell \leq I$ for physical validity.

Therefore, the Mueller matrix elements can now be defined as \cite{GOLDBERG2022185}
\begin{equation}
\begin{aligned}
M_{jk}
&= \frac{1}{2}\sum_\ell
   \operatorname{Tr}\left(
   \alpha_j E_\ell\alpha_k E_\ell^\dagger
   \right).
\end{aligned}
\end{equation}

\subsection{Deriving a quantum operation from a Mueller matrix}
\label{App:Deriving_Quantum Channel_Mueller}
Equation~\eqref{eq:muller_qunatum} shows that each Mueller-matrix element
is the projection of the transformed basis operator
$\mathcal{I}(\alpha_k)$ onto the basis operator $\alpha_j$.
Consequently, the $k$th column of the Mueller matrix contains the
coefficients required to express $\mathcal{I}(\alpha_k)$ in the chosen
operator basis.

To show this explicitly, the action of the quantum operation on
$\alpha_k$ can be shown as
\begin{equation}
\mathcal{I}(\alpha_k)
=
\sum_{m=0}^{3} C_{mk}\alpha_m,
\end{equation}
where the coefficients $C_{mk}$ are initially unknown. Expanding the sum,
\begin{equation}
\mathcal{I}(\alpha_k)
=
C_{0k}I
+
C_{1k}\sigma_z
+
C_{2k}\sigma_x
+
C_{3k}\sigma_y.
\label{eq:channel_expansion}
\end{equation}

Substituting Eq.~\eqref{eq:channel_expansion} into
Eq.~\eqref{eq:muller_qunatum} and using the linearity of the trace gives

\begin{align}
M_{jk}
=
\frac{1}{2}
\Big[
C_{0k}\operatorname{Tr}(\alpha_j I)
+
C_{1k}\operatorname{Tr}(\alpha_j\sigma_z)
\nonumber
&+
C_{2k}\operatorname{Tr}(\alpha_j\sigma_x)
+
C_{3k}\operatorname{Tr}(\alpha_j\sigma_y)
\Big].
\end{align}

As the identity and Pauli operators form an orthogonal basis under the
Hilbert--Schmidt inner product,
$\operatorname{Tr}(\alpha_j\alpha_m)=2\delta_{jm},$ where
$\delta_{jm}= \begin{cases} 1, & j=m,\\ 0, & j\neq m\end{cases}$, all terms therefore vanish except the one for which $j=m$. Thus,

\begin{equation}
C_{jk}=M_{jk}.
\end{equation}

The coefficients appearing in the operator expansion of the quantum operation are therefore exactly the elements of the Mueller matrix.
Substituting these coefficients back into
Eq.~\eqref{eq:channel_expansion} gives
\begin{equation}
\mathcal{I}(\alpha_k)=M_{0k}I+M_{1k}\sigma_z+M_{2k}\sigma_x+M_{3k}\sigma_y.
\end{equation}

More compactly,
\begin{equation}
\
\mathcal{I}(\alpha_k)
=
\sum_{j=0}^{3}
M_{jk}\alpha_j
.
\label{eq:channel_from_mueller}
\end{equation}

Thus, each column of the Mueller matrix specifies the action of the optical operation on one element of the operator basis. Explicitly,

\begin{equation}
\begin{aligned}
\mathcal{I}(I)
&=
M_{00}I
+
M_{10}\sigma_z
+
M_{20}\sigma_x
+
M_{30}\sigma_y,
\\
\mathcal{I}(\sigma_z)
&=
M_{01}I
+
M_{11}\sigma_z
+
M_{21}\sigma_x
+
M_{31}\sigma_y,
\\
\mathcal{I}(\sigma_x)
&=
M_{02}I
+
M_{12}\sigma_z
+
M_{22}\sigma_x
+
M_{32}\sigma_y,
\\
\mathcal{I}(\sigma_y)
&=
M_{03}I
+
M_{13}\sigma_z
+
M_{23}\sigma_x
+
M_{33}\sigma_y.
\end{aligned}
\label{channel_expand}
\end{equation}

In \mbox{Eq.~\eqref{eq:rho_prime}}, the output state $\rho'$ was expressed as the result of a quantum operation acting on the input density matrix $\rho$. Substituting Eq.~\eqref{eq:channel_from_mueller}, which gives the operation's action in terms of the Mueller matrix, gives

\begin{equation}
\begin{aligned}
\rho'
&=
\frac{1}{2}
\sum_{k=0}^{3}
S_k
\left(
\sum_{m=0}^{3}
M_{mk}\alpha_m
\right)
\\
&=
\frac{1}{2}
\sum_{m=0}^{3}
\sum_{k=0}^{3}
M_{mk}
S_k
\alpha_m.
\end{aligned}
\label{eq:rho_prime_1}
\end{equation}

Substituting Eq.~\eqref{eq:rho_prime_1} into Eq.~\eqref{eq:out_stokes_dense}, which defines the output Stokes parameters corresponding to $\rho'$, gives

\begin{equation}
\begin{aligned}
S_j^{\mathrm{out}}
&=
\operatorname{Tr}
\left(
\left(
\frac{1}{2}
\sum_{m,k=0}^{3}
M_{mk}
S_k
\alpha_m
\right)
\alpha_j
\right)
\\
&=
\frac{1}{2}
\operatorname{Tr}
\left(
\sum_{m,k=0}^{3}
M_{mk}
S_k
\alpha_m\alpha_j
\right).
\end{aligned}
\end{equation}

Using the linearity of the trace,

\begin{equation}
S_j^{'}
=
\frac{1}{2}
\sum_{m,k=0}^{3}
M_{mk}
S_k
\operatorname{Tr}
\left(
\alpha_m\alpha_j
\right).
\end{equation}

with the basis operators satisfying the orthogonality relation
$\operatorname{Tr}\left(\alpha_m\alpha_j\right)=2\delta_{mj},$
where $\delta_{mj}$ is the Kronecker delta. Therefore,
\begin{align}
S_j'
&=
\sum_{m,k=0}^{3}
M_{mk}
S_k
\delta_{mj}.
\end{align}

where all terms in the sum of $m$ vanish except the term for which $m=j$ since $
\delta_{mj}
=
\begin{cases}
1, & m=j,\\
0, & m\neq j,
\end{cases}
$.

Hence,
\begin{equation}
S_j'
=
\sum_{k=0}^{3}
M_{jk}
S_k.
\end{equation}

Writing the four output Stokes parameters explicitly,

\begin{equation}
\begin{aligned}
S_0'
&=
M_{00}S_0
+
M_{01}S_1
+
M_{02}S_2
+
M_{03}S_3,
\\
S_1'
&=
M_{10}S_0
+
M_{11}S_1
+
M_{12}S_2
+
M_{13}S_3,
\\
S_2'
&=
M_{20}S_0
+
M_{21}S_1
+
M_{22}S_2
+
M_{23}S_3,
\\
S_3'
&=
M_{30}S_0
+
M_{31}S_1
+
M_{32}S_2
+
M_{33}S_3.
\end{aligned}
\end{equation}

This can be written in matrix form as
\begin{equation}
\begin{pmatrix}
S_0'\\
S_1'\\
S_2'\\
S_3'
\end{pmatrix}
=
\begin{pmatrix}
M_{00} & M_{01} & M_{02} & M_{03}\\
M_{10} & M_{11} & M_{12} & M_{13}\\
M_{20} & M_{21} & M_{22} & M_{23}\\
M_{30} & M_{31} & M_{32} & M_{33}
\end{pmatrix}
\begin{pmatrix}
S_0\\
S_1\\
S_2\\
S_3
\end{pmatrix}.
\end{equation}

Therefore, using Eq.~\ref{eq:channel_from_mueller}, we can retrieve 
$\mathbf{S}_{\mathrm{out}}=M\mathbf{S}_{\mathrm{in}}.$

\subsection{Physical validation}
\label{Physical_val}
\subsubsection{Complete positivity}
The relation in Eq.~\ref{eq:channel_from_mueller} determines the action of the map on a complete
operator basis and hence specifies the linear map uniquely.
An arbitrary real $4\times4$ matrix, however, need not define
a physical quantum operation. The reconstructed map must be
completely positive and trace-non-increasing to describe the
passive, lossy process considered in the protocol in this work.

To determine whether the reconstructed linear map \(\mathcal I\) defines a completely positive quantum operation, we start by constructing its Choi matrix. For a map acting on a two dimensional Hilbert space, the Choi matrix is defined as
\begin{equation}
\mathcal{C}_{\mathcal I}
=
\sum_{m,n=0}^{1}
\ket{m}\bra{n}
\otimes
\mathcal I\!\left(\ket{m}\bra{n}\right).
\label{eq:choi-definition}
\end{equation}

In the ordered computational basis, \(\{\ket{0},\ket{1}\}\) of the input space, the Choi matrix can be written as a \(2\times2\) block matrix, where each block is itself a \(2\times2\) matrix, giving \(\mathcal{C}_{\mathcal I}\) as a \(4\times4\) matrix. We can therefore write Eq.~\eqref{eq:choi-definition} as
\begin{equation}
\mathcal{C}_{\mathcal I}
=
\begin{pmatrix}
\mathcal I(\ket{0}\bra{0})
&
\mathcal I(\ket{0}\bra{1})
\\[4pt]
\mathcal I(\ket{1}\bra{0})
&
\mathcal I(\ket{1}\bra{1})
\end{pmatrix},
\label{eq:choi-block-general}
\end{equation}

As expanded upon in Eq.~\ref{channel_expand}, we label the action of \(\mathcal I\) on the Pauli basis as
\begin{equation}
A=\mathcal I(I),\qquad
B=\mathcal I(\sigma_z),\qquad
C=\mathcal I(\sigma_x),\qquad
D=\mathcal I(\sigma_y).
\label{eq:map-pauli-action}
\end{equation}
with the ordering corresponding to the Stokes convention, 
$(S_0,S_1,S_2,S_3) \longleftrightarrow
(I,\sigma_z,\sigma_x,\sigma_y).$

The computational basis can be mapped to the Pauli basis with the following relation:

\begin{equation}
\begin{aligned}
\ket{0}\bra{0}
&=
\frac{I+\sigma_z}{2},
&
\ket{0}\bra{1}
&=
\frac{\sigma_x+i\sigma_y}{2},
\\[4pt]
\ket{1}\bra{0}
&=
\frac{\sigma_x-i\sigma_y}{2},
&
\ket{1}\bra{1}
&=
\frac{I-\sigma_z}{2}.
\end{aligned}
\label{eq:computational-pauli-expansion}
\end{equation}

Using the linearity of \(\mathcal I\), together with Eq.~\eqref{eq:map-pauli-action}, gives
\begin{equation}
\begin{aligned}
\mathcal I(\ket{0}\bra{0})
&=
\frac{1}{2}(A+B),
&
\mathcal I(\ket{0}\bra{1})
&=
\frac{1}{2}(C+iD),
\\[4pt]
\mathcal I(\ket{1}\bra{0})
&=
\frac{1}{2}(C-iD),
&
\mathcal I(\ket{1}\bra{1})
&=
\frac{1}{2}(A-B).
\end{aligned}
\label{eq:map-computational-action}
\end{equation}

Substituting these expressions into Eq.~\eqref{eq:choi-block-general}, we obtain the resulting $4\times4$ Choi matrix, with each entry itself being a $2\times2$ matrix, as

\begin{equation}
\mathcal{C}_{\mathcal I}
=
\frac{1}{2}
\begin{pmatrix}
A+B & C+iD\\[4pt]
C-iD & A-B
\end{pmatrix}.
\label{eq:choi-pauli-block}
\end{equation}

For the map $\mathcal{I}$ to be completely positive, the Choi matrix must
be positive semidefinite,
\begin{equation}
C_{\mathcal{I}} \succeq 0.
\end{equation}

Equivalently,
\begin{equation}
\mathbf{v}^{\dagger}
C_{\mathcal{I}}
\mathbf{v}
\geq 0,
\qquad
\forall\,\mathbf{v}\in\mathbb{C}^{4}.
\end{equation}

This condition is equivalent to requiring all eigenvalues of the Choi
matrix to be non-negative,
\begin{equation}
\lambda_{\ell}(C_{\mathcal{I}})\geq0
\qquad
\text{for all } \ell.
\end{equation}

The spectral decomposition of the Choi matrix can therefore be written as
\begin{equation}
C_{\mathcal{I}}
=
\sum_{\ell}
\lambda_{\ell}
\ket{v_{\ell}}
\bra{v_{\ell}},
\qquad
\lambda_{\ell}\geq0,
\end{equation}
where the eigenvectors are written as, with the first index labelling the input basis state $\ket{m}$ and the second the output basis state, following the block structure of \mbox{Eq.~\eqref{eq:choi-block-general}},
\begin{equation}
\ket{v_{\ell}}
=
\begin{pmatrix}
v_{\ell,00} \\
v_{\ell,01} \\
v_{\ell,10} \\
v_{\ell,11}
\end{pmatrix}.
\end{equation}

Each eigenvector can then be reshaped into a $2\times2$ operator. The
corresponding Kraus operator is defined as
\begin{equation}
E_{\ell}
=
\sqrt{\lambda_{\ell}}
\begin{pmatrix}
v_{\ell,00} & v_{\ell,10} \\
v_{\ell,01} & v_{\ell,11}
\end{pmatrix}.
\end{equation}

The quantum map reconstructed from the Choi matrix can therefore be
written in Kraus form as
\begin{equation}
\mathcal{I}(\rho)
=\sum_{\ell}E_{\ell}\rho E_{\ell}^{\dagger}.
\end{equation}

\subsubsection{Trace preservation and trace non-increase}
In order to determine whether a map is trace-preserving or trace-non-increasing, we take the partial trace of the Choi matrix in Eq.~\ref{eq:choi-pauli-block} over the output Hilbert space, 
$T=\operatorname{Tr_{partial}} (\mathcal{C}_{\mathcal I})$.

The partial trace is
\begin{equation}
T
=
\frac{1}{2}
\begin{pmatrix}
\operatorname{Tr}(A+B)
&
\operatorname{Tr}(C+iD)
\\[4pt]
\operatorname{Tr}(C-iD)
&
\operatorname{Tr}(A-B)
\end{pmatrix}.
\label{eq:partial-trace-blocks}
\end{equation}

Using \mbox{Eqs.~\eqref{channel_expand}} and \mbox{\eqref{eq:map-pauli-action}} and the established ordering $(\alpha_0,\alpha_1,\alpha_2,\alpha_3) =
(I,\sigma_z,\sigma_x,\sigma_y)$, we get
\begin{equation}
\begin{aligned}
\frac{1}{2}\operatorname{Tr}(A+B)
&=
M_{00}+M_{01},
&
\frac{1}{2}\operatorname{Tr}(C+iD)
&=
M_{02}+iM_{03},
\\[4pt]
\frac{1}{2}\operatorname{Tr}(C-iD)
&=
M_{02}-iM_{03},
&
\frac{1}{2}\operatorname{Tr}(A-B)
&=
M_{00}-M_{01}.
\end{aligned}
\label{eq:trace-blocks}
\end{equation}

Substituting these expressions into Eq.~\eqref{eq:partial-trace-blocks} gives
\begin{equation}
T
=
\begin{pmatrix}
M_{00}+M_{01}
&
M_{02}+iM_{03}
\\[4pt]
M_{02}-iM_{03}
&
M_{00}-M_{01}
\end{pmatrix}.
\label{eq:T-Mueller}
\end{equation}

Equivalently we can write Eq.~\mbox{\eqref{eq:T-Mueller}} as,
\begin{equation}
T
=
M_{00}I
+
M_{01}\sigma_z
+
M_{02}\sigma_x
-
M_{03}\sigma_y.
\label{eq:T-Pauli}
\end{equation}
For a completely positive map to also be trace-non-increasing, its partial trace $T$ must fulfil the condition
\begin{equation}
0\le T\le I.
\label{eq:TNI-condition}
\end{equation}
The lower bound follows from \(\mathcal{C}_{\mathcal I}\ge0\) which comes from the relation that the partial trace of any positive semidefinite matrix is always positive semidefinite, while the upper bound can be written as
\begin{equation}
I-T\ge0.
\end{equation}

Let us consider the characteristic equation to express this condition in terms of the first row of the Mueller matrix, which determines the output intensity $S'_0$, 
\begin{equation}
\det(T-zI)=0.
\end{equation}
Using Eq.~\eqref{eq:T-Mueller} we can expand,
\begin{align}
\det(T-zI)
&=
(M_{00}+M_{01}-z)
(M_{00}-M_{01}-z)
\nonumber\\
&\quad
-
(M_{02}+iM_{03})
(M_{02}-iM_{03})
\nonumber\\
&=
(M_{00}-z)^2
-
M_{01}^2
-
M_{02}^2
-
M_{03}^2.
\end{align}

We can define $R=\sqrt{M_{01}^2+M_{02}^2+M_{03}^2}$ giving the two eigenvalues of $T$ as $z_-=M_{00}-R$ and $z_+=M_{00}+R.$

The condition \(0\preceq T\preceq I\) is therefore equivalent to
\begin{equation}
0\leq M_{00}-R\leq M_{00}+R\leq 1.
\label{eq:Mueller-TNI-condition}
\end{equation}

For a normalised input Stokes vector, the first row of the Mueller matrix determines the output intensity and thus, physically, Eq.~\eqref{eq:Mueller-TNI-condition} acts as the fundamental bound ensuring no physical input state yields an output probability outside the interval $[0,1]$. This is tied to the fact that the extremal eigenvalues of $T$ correspond to the minimum and maximum possible output intensities over all fully polarised inputs.

Equivalently, we can deduce
\begin{equation}
R\leq M_{00},
\qquad
M_{00}+R\leq1.
\end{equation}

In the strict case of a trace-preserving map, the equality
\begin{equation}
T=I
\end{equation}
must be satisfied. Imposing this constraint on the Mueller matrix would require the elements of the first row to yield
$M_{00}=1$ and $M_{01}=M_{02}=M_{03}=0.$ This gives the full trace of the Choi matrix as $\operatorname{Tr}(\mathcal{C}_{\mathcal I}) = 2M_{00}$ which agrees with a trace-preserving qubit map with trace $\operatorname{Tr}(\mathcal{C}_{\mathcal I}) = 2$. While the full trace of the Choi matrix serves as a necessary scalar condition, it is not sufficient to ensure that the quantum map is trace-preserving. Suppose, for example, if we consider a map parametrised by $M_{00}=1$ and $M_{01}=1$ (with $M_{02}=M_{03}=0$), it will have partial trace of
$T = \begin{pmatrix} 2 & 0  \\ 0 & 0 \end{pmatrix}$, which has a full trace of $\operatorname{Tr}(T)=2$. Although this perfectly satisfies the scalar condition, the map is not trace-preserving. It corresponds to an unphysical process that amplifies one state to 200$\%$ probability while completely annihilating the orthogonal state. Thus, true trace preservation strictly requires the full operator condition $T=I$. 

\section{Simulation Results without a Sample}\label{App:NoSample}

In the absence of a sample, and with two beam splitter interactions per cycle with each contributing an angle $\pi/4n$, where $n$ is the number of interferometer cycles, the total rotation after $n$ cycles is 

\begin{equation}
     2n\frac{\pi}{4n}=\frac{\pi}{2}.
\end{equation}

So if the photon started in path~2, it would transfer to path 1, leading to detection at ICCD0. Thus, whatever the input polarisation, the output polarisation equals the input polarisation; the simulation confirms $P(\mathrm{ICCD0})=1$ for all six inputs at $n=100$ (the seven reference parameter sets play no role once the sample is removed), with absorption below $10^{-13}$. 

\section{Modelling Detection/Interaction Rates for the Full Model}\label{App:DetectionRates}
\begin{figure}[htbp]
    \centering
    \includegraphics[width=0.59\linewidth]{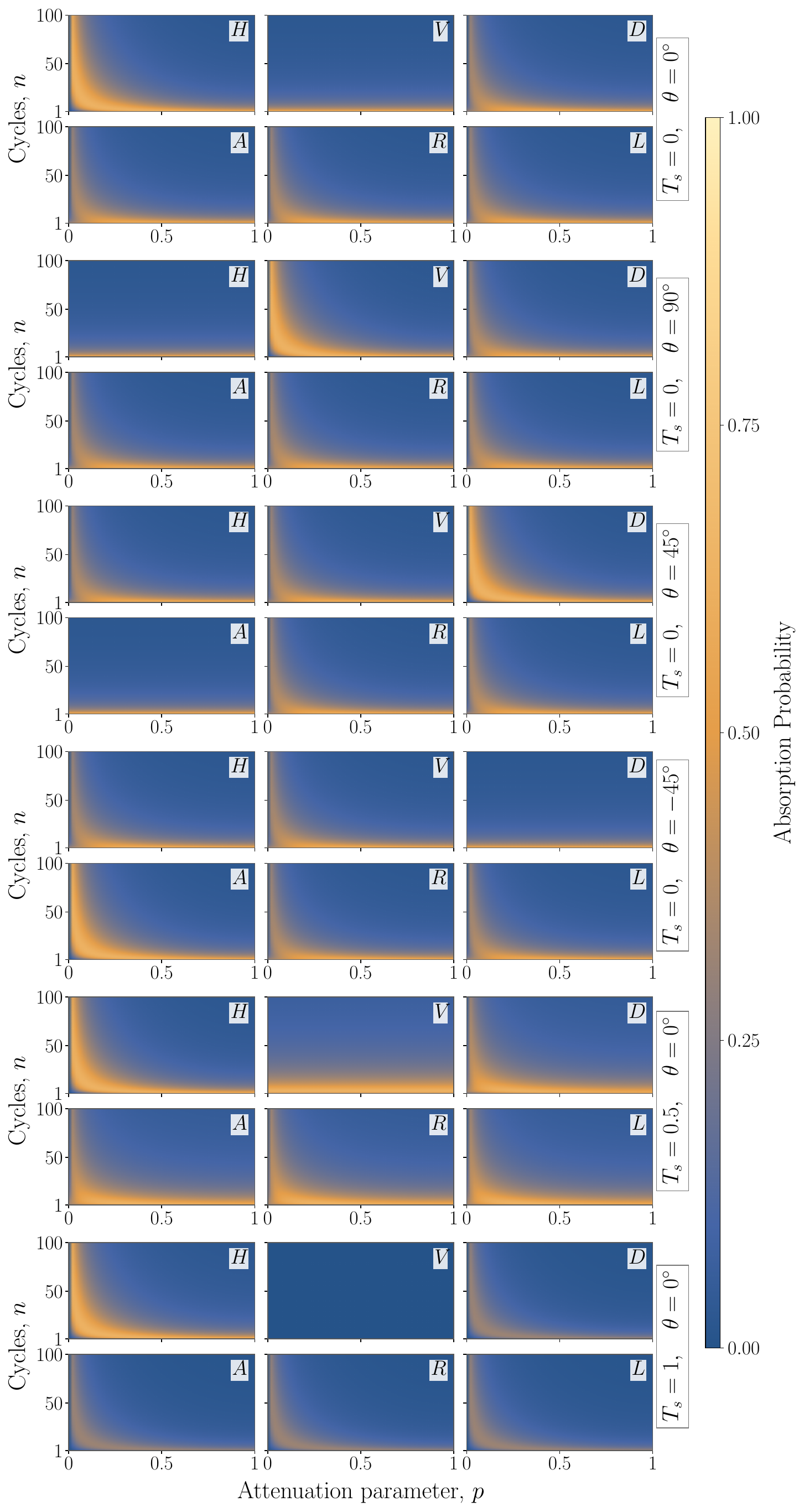}
    \caption{Probability of the sample absorbing the probe photon in an overall run of the protocol, as a function of $p$, the attenuation parameter, and $n$, the number of interferometer cycles in the run. The attenuation parameter $p$ sets the single-pass principal transmittance $T_f=(1-p)^2$. Each $2\times3$ block corresponds to one sample, and the small box in each panel indicates the input polarisation. From top to bottom the blocks have $T_s=0$ with $\theta=0^\circ$, $90^\circ$, $45^\circ$ and $-45^\circ$ (the linear polarisers of Fig.~\mbox{\ref{fig.det-abs}} at $p=0$; the $0\%$ filter coincides with the first block), followed by $T_s=0.5$ and $T_s=1$ with $\theta=0^\circ$. The principal transmittance, $T_s$, and the reference-axis angle, $\theta$, are specified for each sample while $T_f$ changes with $p$.}
    \label{fig.absorption_compact}
\end{figure}

\begin{figure}[htbp]
    \centering
    \includegraphics[width=0.59\linewidth]{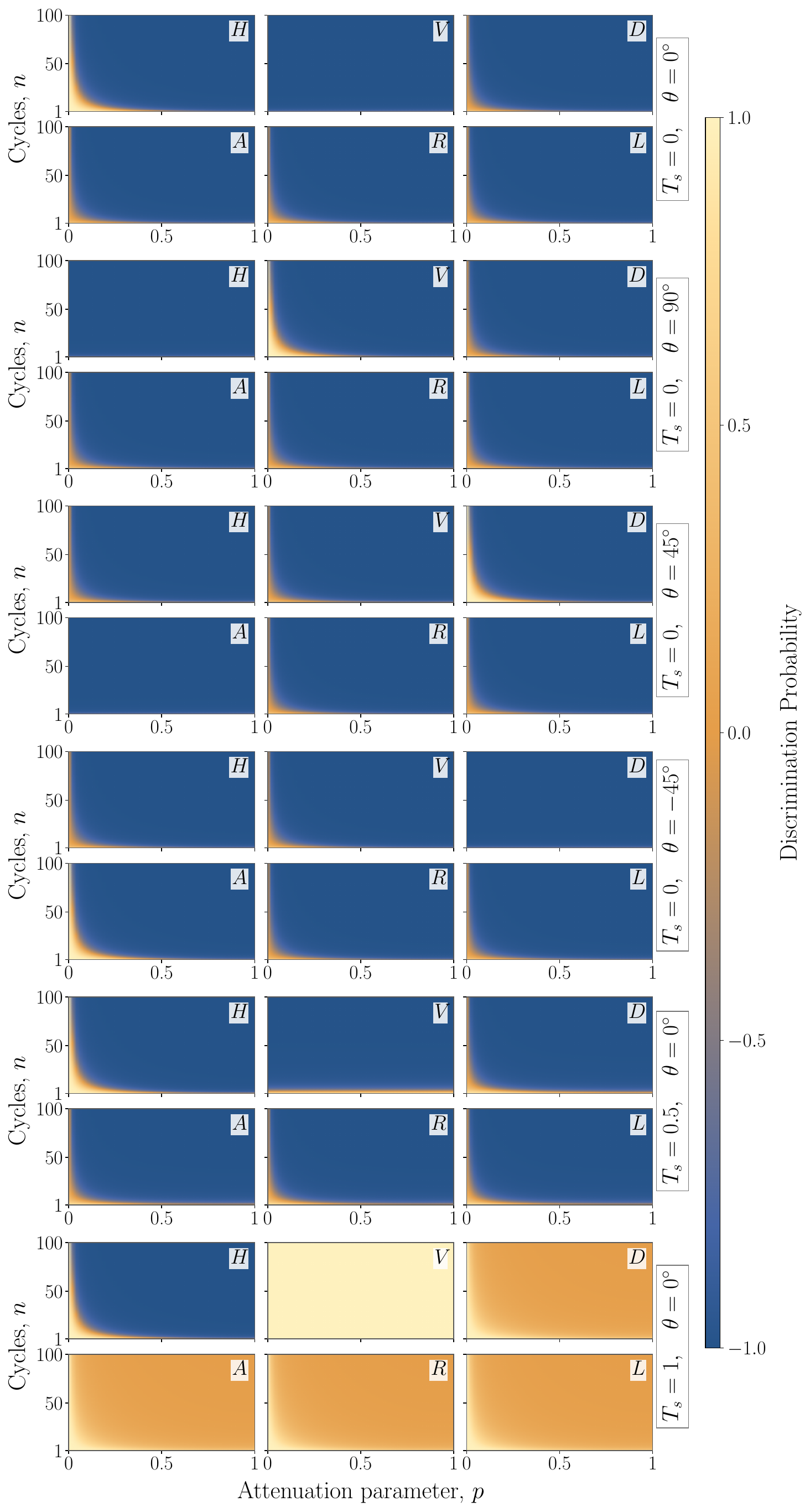}
    \caption{Port asymmetry $[P(\mathrm{ICCD0})-P(\mathrm{ICCD1})]/[P(\mathrm{ICCD0})+P(\mathrm{ICCD1})]$ (labelled discrimination in the figure), with the sign convention of Fig.~\mbox{\ref{fig.dis}} (a transmitted input gives $+1$), as a function of $p$, the attenuation parameter, and $n$, the number of interferometer cycles in the run. The attenuation parameter $p$ sets the single-pass principal transmittance $T_f=(1-p)^2$. Each $2\times3$ block corresponds to one sample, and the small box in each panel indicates the input polarisation. From top to bottom the blocks have $T_s=0$ with $\theta=0^\circ$, $90^\circ$, $45^\circ$ and $-45^\circ$, followed by $T_s=0.5$ and $T_s=1$ with $\theta=0^\circ$. The principal transmittance, $T_s$, and the reference-axis angle, $\theta$, are specified for each sample while $T_f$ changes with $p$.}
    \label{fig.discrimination_compact}
\end{figure}

Figs.~\mbox{\ref{fig.absorption_compact}} and \mbox{\ref{fig.discrimination_compact}} give, for the reference samples of \mbox{Fig.~\ref{fig.det-abs}} and each input polarisation, the absorption probability and the port asymmetry of the full model of Sec.~\mbox{\ref{sec:results}} as the transmittance along the fast axis is swept, $T_f=(1-p)^2$; they are the full-model counterparts of Figs.~\mbox{\ref{fig:TotAbsorbed}} and \mbox{\ref{fig:Discrim}}.

\section{Signal-to-Noise Ratio of the Protocol for Each Reference Sample}\label{App:SNR}
\begin{figure}[htbp]
    \centering
    \includegraphics[width=0.75\linewidth]{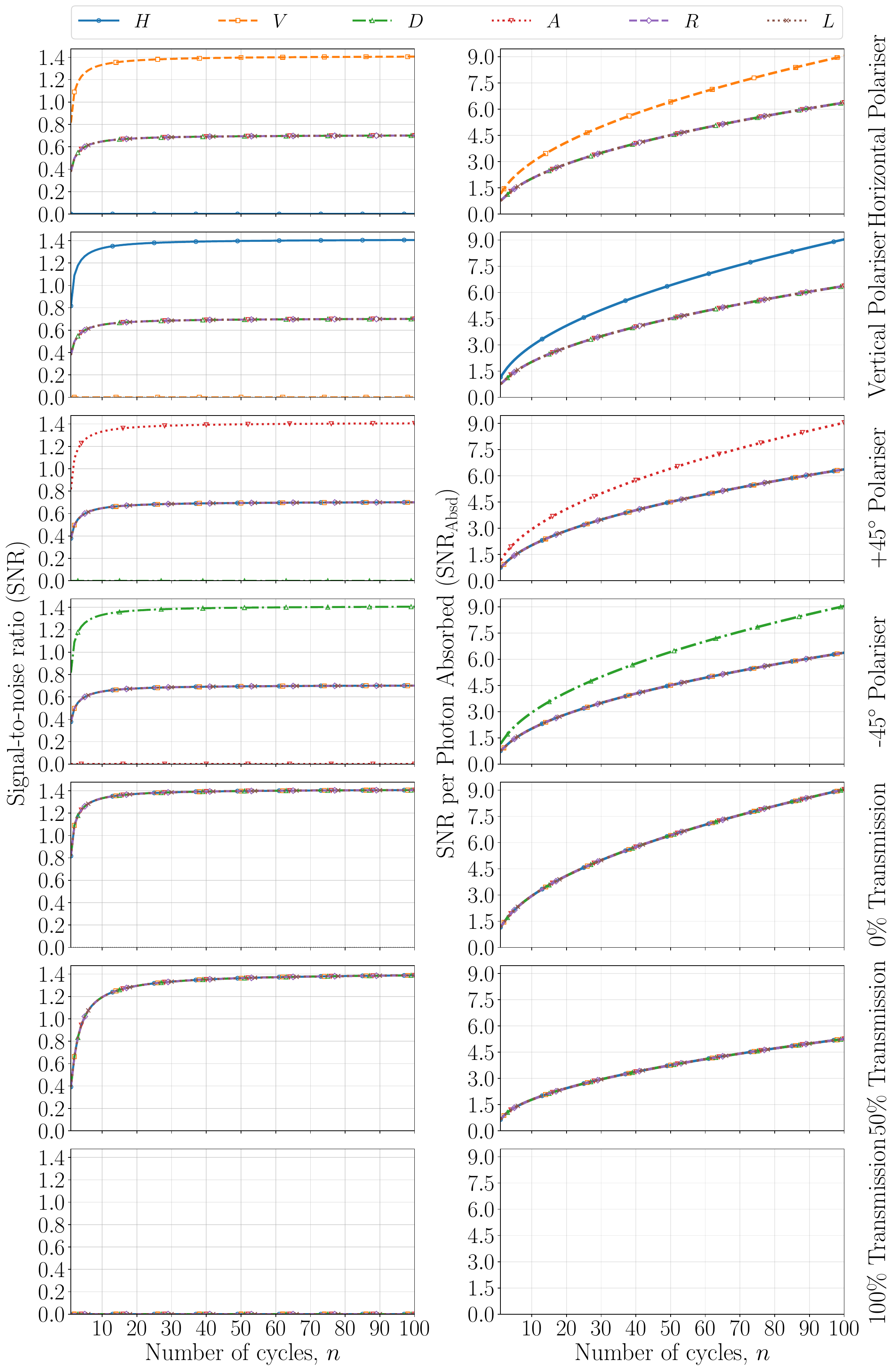}
    \caption{Left: SNR for distinguishing the sample-present from the sample-absent case for each polarisation-pixel mode, normalised by dividing by $\sqrt{\bar{N}}$, the square root of the average number of photons emitted into that mode. This is a function of $n$, the number of interferometer cycles in the run. Right: the same SNR per absorbed photon, $\mathrm{SNR}/\sqrt{P_{\mathrm{abs}}}$, for each polarisation-pixel mode, also as a function of $n$. Each row is one of the samples of \mbox{Fig.~\ref{fig.det-abs}}, and each line is one of the six input polarisation states $H$, $V$, $D$, $A$, $R$ and $L$.}
    \label{fig:snr_comparison}
\end{figure}

Fig.~\mbox{\ref{fig:snr_comparison}} shows, for each reference sample of \mbox{Fig.~\ref{fig.det-abs}} and each input polarisation, the SNR for distinguishing the sample-present from the sample-absent case (left), which vanishes when the input is aligned with a fully transmitting axis, and the same quantity divided by the square root of the absorbed fraction (right). The right panel is the per-sample form of the quantity plotted in \mbox{Fig.~\ref{fig:SNRPerAbsd}}, and reaches about $9$ at $n=100$ for an opaque mode, with values for a horizontal polariser (sample) given in Table~\ref{tab:snr_horizontal_opaque} with the other linear polarisers giving equivalent values for input states with the same alignment relative to their transmitting axes.

\begin{table}[h]
\centering
\begin{tabular}{l|r|r|r}
\toprule
Input &  $P_\mathrm{Absd}$ &  $\mathrm{SNR}$ &  $\mathrm{SNR} / \sqrt{N_\mathrm{Absd}}$ \\
\midrule
\hline
    H &   - &       - &      - \\
 V &    0.024 &       1.406 &        9.037 \\
     D &     0.012 &       0.701 &   6.370\\
    A &     0.012 &       0.701 &   6.370 \\
    R &     0.012 &       0.701 &   6.370 \\
     L &     0.012 &       0.701 &  6.370 \\
\hline

\bottomrule
\end{tabular}
\caption{Results at $n=100$ for a horizontal polariser, showing the probability of absorption, SNR, and SNR  per absorbed photon corresponding to each input polarisation state.}
\label{tab:snr_horizontal_opaque}
\end{table}

\end{document}